\documentclass[letterpaper]{article}

\usepackage[nonatbib,preprint]{neurips_2023}
\usepackage[numbers]{natbib}

\usepackage[utf8]{inputenc} % allow utf-8 input
\usepackage[T1]{fontenc}    % use 8-bit T1 fonts
\usepackage{hyperref}       % hyperlinks
\usepackage{url}            % simple URL typesetting
\usepackage{booktabs}       % professional-quality tables
\usepackage{amsfonts}       % blackboard math symbols
\usepackage{nicefrac}       % compact symbols for 1/2, etc.
\usepackage{microtype}      % microtypography
\usepackage{xcolor}         % colors
\usepackage{graphicx}
\usepackage[]{algorithm2e}
\usepackage{amsmath}
\usepackage{mathtools}      % for coloneqq symbol
\usepackage{multirow}       % for multiple rows in a single table cell
\usepackage{enumitem }      % to reduce indent on lists
\usepackage{longtable}      % for long tables
\usepackage{subcaption}     % for subfigures and captions

\title{Grid-SiPhyR: An end-to-end learning to optimize framework for combinatorial problems in power systems}

\author{%
  Rabab Haider \\
  Department of Mechanical Engineering\\
  Massachusetts Institute of Technology\\
  Cambridge, MA 02149 \\
  \texttt{rhaider@mit.edu} \\
  \And
  Anuradha M. Annaswamy \\
  Department of Mechanical Engineering\\
  Massachusetts Institute of Technology\\
  Cambridge, MA 02149 \\
  \texttt{aanna@mit.edu} \\
}

\begin{document}

\maketitle

\begin{abstract}
    Mixed integer problems are ubiquitous in decision making, from discrete device settings and design parameters, unit production, and on/off or yes/no decision in switches, routing, and social networks. Despite their prevalence, classical optimization approaches for combinatorial optimization remain prohibitively slow for fast and accurate decision making in dynamic and safety-critical environments with hard constraints. To address this gap, we propose Si\textbf{PhyR} (\textit{pronounced: cipher}), a physics-informed machine learning framework for end-to-end learning to optimize for combinatorial problems. Si\textbf{PhyR} employs a novel physics-informed rounding approach to tackle the challenge of combinatorial optimization within a differentiable framework that has certified satisfiability of safety-critical constraints. We demonstrate the effectiveness of Si\textbf{PhyR} on an emerging paradigm for clean energy systems: dynamic reconfiguration, where the topology of the electric grid and power flow are optimized so as to maintain a safe and reliable power grid in the presence of intermittent renewable generation. Offline training of the unsupervised framework on representative load and generation data makes dynamic decision making via the online application of Grid-Si\textbf{PhyR} computationally feasible.
\end{abstract}

\section{Introduction}
Exactly solving combinatorial optimization problems (also called mixed integer programs, MIP) is an NP-hard problem that requires exponential time to solve. Traditional optimization methods become intractable especially when requiring dynamic decisions (fast + frequent) and high degree of accuracy to remain within operating constraints or design limits. Traditional solvers also struggle to take advantage of the structures present in repeatedly solving an optimization problem, and warm-start techniques may struggle when parameters vary rapidly -- such as with solar generation forecasts or design under time-varying conditions. Machine learning (ML) algorithms are natural candidates as function approximators, to learn the underlying problem structure of complex, high-dimensional, and nonlinear optimization spaces. However, out-of-the-box implementations typically cannot enforce hard constraints or address mixed integer variables.

In this work we introduce Si\textbf{PhyR} (\textit{Si}gmoidal \textit{Phy}sics-Informed \textit{R}ounding; \textit{pronounced as: `cipher'}), a physics-informed ML framework for end-to-end learning to optimize for combinatorial problems. We embed within a differentiable framework the physics equations governing the behaviour of an arbitrary system in the form of equality and inequality constraints. To tackle the challenge of mixed integer variables, we introduce a general physics-informed rounding approach and the concept of a rounding function which is defined by a set of physics equations of the underlying problem. The physics-informed approach guarantees certified satisfiability of all equality constraints, a subset of inequality constraints, and integer variable constraints. Notably, our Si\textbf{PhyR} framework is an unsupervised method that accomplishes an end-to-end learning by directly integrating the task of optimization within the ML framework, without requiring access to the optimal solution of each training instance and replaces traditional optimization solvers.

% \textcolor{red}{Notably, our Si\textbf{PhyR} framework is an unsupervised method which does not depend on access to the optimal solution of each training instance, which can be burdensome and computationally prohibitive to obtain. Instead our physics-informed approach enables the predictor to \textit{learn to optimize} and provide good and fast solutions which preserve feasibility of key physical constraints.}

Our key contributions are: \textbf{(1) A framework for end-to-end learning for combinatorial problems:} We describe a general framework, Si\textbf{PhyR}, with a physics-informed rounding approach which leverages physics-based system constraints to guide the neural predictor towards feasible and good solutions for dynamic decision making (fast + frequent) in the presence of mixed integer variables. \textbf{(2) Implementation of SiPhyR on a real-world problem for clean energy systems:} We develop Grid-Si\textbf{PhyR}, an implementation of our end-to-end framework for dynamic grid reconfiguration, to optimize the topology of and power flows on the electric grid in the presence of intermittent renewable generation. \textbf{(3) Demonstration of the effectiveness of Grid-SiPhyR:} We conduct experiments on two canonical distribution grids and show that Grid-Si\textbf{PhyR} can learn to optimize while preserving feasibility and enjoying scalability. We also develop representative datasets for power systems researchers to support algorithm development and testing in the absence of real-world data.

\section{Related work}
% \paragraph{ML for Optimization} 
\textbf{ML for Optimization.} The need for fast and repeated solutions to optimization problems has pushed for ML-based solutions for directly optimizing within a neural framework \cite{donti_DC3}, learning new algorithms for optimization \cite{ML_for_algorithm_design}, and improving the performance of existing solvers. The latter encompasses the vast majority of efforts, including hyperparameter optimization \cite{Maclaurin_hyperopt_learning}; identifying active constraint sets \cite{Misra_ML_for_activeconstraints,Deka_dcopf_activesets_ML}; learning warm start techniques \cite{Baker_2019warmstart,Xavier_SCUC}; and mapping strategies used by optimization solvers \cite{Bertsimas_Voice_2021}. Approaches specifically for MIPs include the design of primal heuristics \cite{Wang_2022_heuristicsMILPs, Shen_IEEE_heuristicsMIPs}, including neural diving \cite{Nair_MIPNN_2020} and neural branching \cite{Gasse_COGNN_2019,Khalil_MLbranching}. Recent surveys on ML for MIPs includes a review of variable branch selection, cutting plane methods, and heuristics such as feasibility pump algorithms \cite{Zhang_MIPML_survey}; and using graph neural networks (GNN) directly as solvers or to enhance exact solvers \cite{Khalil_COGNN_2021}. Our approach falls in the first group: directly solving the MIP using an unsupervised neural network within an end-to-end learning to optimize framework. We propose the used of physics-informed ML to significantly improve prediction performance. % These surveys and references within present a detailed report of the state of art. 

% \paragraph{ML for Power Systems} 
\textbf{ML for Power Systems.} There is a growing body of literature employing ML techniques for power systems tasks including optimal power flow (OPF) \cite{deepopf, deepopfplus, donti_DC3, Zhou_DRL_2020, Dobbe_decentralizedML,Fioretto_lagrangeduals,Zamzam_SmartGridComm2020,vpoor_opf_topologyreconfig}, probabilistic power flow \cite{baosen_probpf}, security constrained unit commitment \cite{Martinez_securityboundary,deepopf_SCUC,Xavier_SCUC}, fault isolation \cite{Li_faultlocation_CNN}, and reconfiguration \cite{Junlakarn_tree, Yin_reconfig, Zhu_linealgoselection, Subramanian_2021}. These works all leverage the significant reduction in computational runtimes of ML-based methods as compared to classical methods such as commercial solvers \cite{deepopf}. We argue that \textbf{physics-informed ML is necessary to enable high prediction accuracy and satisfy critical physics and operating constraints in an explainable manner}. Prior work has incorporated power physics models into neural layers \cite{deepopf,donti_DC3}, embedded into GNN structures \cite{Diehl_2019_warmstartOPFGNN,Liu_2022_GNNACOPF,Donon_2020_powerflowGNN,Li_2022_GridWarm}, and satisfied voltage and generator limits by using Lagrangian duals \cite{Fioretto_lagrangeduals,Spyros_2022_PhyMLACpowerflow}, projections \cite{deepopf,deepopfplus}, or gradient-based algorithms \cite{donti_DC3}. Physics-informed ML has also been used to improve computational performance of weight updates \cite{baosen_probpf}. Our work presents a significant extension to the prior literature towards \textbf{solving mixed-integer problems in a neural framework while satisfying critical physics constraints}.

% Rather than tackling the MIP directly, decision trees and classification methods have been employed for grid reconfiguration, but have shown to underperform compared to other ML methods such as neural networks \cite{Zhu_linealgoselection}. Our work takes a step towards ML for MIPs, using a novel physics-informed rounding heuristic.

% , and algorithm selection for grid reconfiguration \cite{Zhu_linealgoselection}
% \textcolor{red}{Add a sentence on what the ML is not doing -- No integer variables. Have not done x problem which is very imp and done all the time}

% \paragraph{Traditional Methods for Reconfiguration}
\textbf{Traditional Methods for Reconfiguration.} The reconfiguration problem has been extensively studied for outage scenarios where switches are operated to maximally serve loads, and to improve grid efficiency by minimizing line losses \cite{Baran_reconfig_1989}. Classical approaches include single loop optimization \cite{Fan_singleloop_1996} and heuristics for approximating losses without extensive power flow calculations \cite{Civanlar_reconfig_1988}. With increased computing abilities and successful commercial MIP software, literature in the 2000's focused on modeling the power physics of reconfiguration as a convex problem \cite{Taylor_reconfigModel}, with a particular focus on radiality constraints \cite{Lei_radiality_2020, Wang_radiality_2020, Ahmadi_radiality_2015, Lavorato_radiality_2012}. Despite the advances in MIP solvers, reconfiguration for realistic power grids remains computationally intractable \cite{Carvalho_reconfig2021}; instead literature has proposed using various heuristic methods including genetic algorithms, constraint elimination, and repeatedly solving smaller sub-optimization problems \cite{crozier_optimalTransmissionSwitching}. However, these techniques may not provide optimality or feasibility guarantees, and remain computationally prohibitive for dynamic applications. We propose physics-informed ML to enable dynamic decisions for reconfiguration with feasibility guarantees.
% where switches are operated to ensure demand is met without violating line thermal constraints and during contingencies. 

\section{Si\textbf{PhyR}: learning to optimize for combinatorial problems}
\begin{figure}
    \centering
    \includegraphics[width=0.9\linewidth]{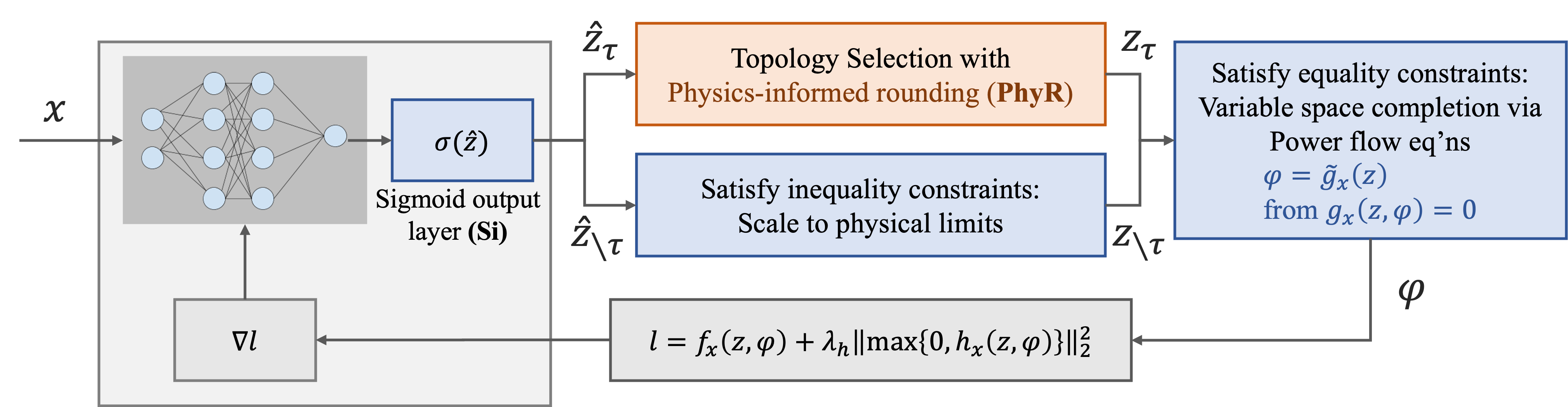}
    \caption{Si\textbf{PhyR}: A physics-informed machine learning framework for end-to-end learning to optimize for a class of combinatorial problems, employing a sigmoidal output layer and a physics-informed algorithm to recover integer solutions}
    % applied to the grid reconfiguration problem to optimally determine grid topology and power dispatch for all generators
    \label{fig:neural_framework}
\end{figure}

We propose Si\textbf{PhyR} (\textit{pronounced: cipher}), a physics-informed machine learning framework for end-to-end learning to optimize for a class of combinatorial problems. Consider a mixed integer problem which given input data $x\in\mathbb{R}^d$ finds $\psi$ that minimizes $f(\psi)$ subject to a set of constraints,  %equality and inequality 
\begin{align} \label{eq:general_MIP}
    \min_{\substack{\psi =[z_\tau, z_{\setminus\tau},\varphi], \\ z_\tau \in \mathbb{Z}_2^m,[z_{\setminus\tau},\varphi]\in\mathbb{R}^n}} f_x(\psi), \quad s.t. \quad g_x(\psi) = 0, \;\; h_x(\psi) \leq 0, \;\; b_x(z_{\setminus\tau}) = 0,
\end{align}
where $f, g, h$ are potentially nonlinear and nonconvex, and $b$ has a particular structure discussed in~\ref{sec:phyR}. Leveraging variable space reduction techniques, the decision variable $\psi$ has been separated into \textit{independent} $z$ and \textit{dependent} variables $\varphi$. Variables $z$ are divided into topological and other variables as $z=[z_\tau, z_{\setminus\tau}]$. Knowledge of $z$ and the function $g$ permits the calculation of $\varphi$, i.e. $\varphi =\tilde{g}_x(z)$. In general, this decomposition is non-unique. It critically depends on the structure of the given problem, which determines the relationship between the sets of variables, and requires domain knowledge to exploit the underlying problem structure to produce good solutions. This optimization problem can be cast as a learning problem with a neural network parameterized by $\theta$ as $\psi = N_\theta(x)$, with suitable accommodations made for the binary variables and constraints. To solve such a problem, we propose an unsupervised neural network shown in Fig.~\ref{fig:neural_framework} composed of five key components described next.

% \paragraph{Lightweight neural network:} 
\textbf{Lightweight neural network:} a neural network with a sigmoidal (Si) output layer. The neural network predicts the \textit{independent variables}, denoted as $\hat{z}$, from the input data $x$. The output of the neural network is divided into two sets: a vector of probabilities describing the likelihood that a binary variable takes on the value of 1, $\hat{z}_\tau=\mathbb{P}(z_\tau = 1)$; and the prediction for the continuous variables $\hat{z}_{\setminus \tau}$. The sigmoidal function is a cumulative distribution function of the logistic distribution, and establishes the interpretation of $\hat{z}$ as a probability. 

% \paragraph{Physics-informed rounding (PhyR):} 
\textbf{Physics-informed rounding (PhyR):} algorithm to recover integer solutions from the vector of probabilities. This layer uses the constraints $b$ to design a \textit{rounding function} $R_b$ that converts probabilities $\hat{z}_\tau$ (input) to binary decisions $z_\tau$ (output). 

% \paragraph{Inequality constraint layers:} 
\textbf{Inequality constraint layers:} the prediction from the neural network $\hat{z}_{\setminus \tau}$ are scaled onto box constraints, using a sigmoidal function mapping. This ensures that inequality constraints pertaining to the independent variables are have certified satisfiability. This layer acts in parallel to the physics-informed rounding. The output of this layer are the scaled variables $z_{\setminus \tau}$.

% \paragraph{Equality constraint layers:} 
\textbf{Equality constraint layers:} leveraging techniques for variable space reduction, the equality constraints are used to calculate the \textit{dependent variables} $\varphi$ from the independent variables $z = \left[ z_\tau \; z_{\setminus \tau}\right]$, as $\varphi=\tilde{g}_x(z)$. This ensures that equality constraints have certified satisfiability.

% \paragraph{Loss function:} 
\textbf{Loss function:} the neural network \textit{learns to optimize} by using an unsupervised framework. The loss function is composed of the objective function of the MIP in \eqref{eq:general_MIP} and regularization via a soft-loss penalty for violating physical constraints (i.e. inequality constraints of dependent variables). The resulting loss function is $l = f_x(\psi) + \lambda_h \Vert \text{max}\left\{0, h_x(z,\varphi)\right\} \Vert_2^2$, where $\lambda_h$ is the soft-loss hyperparameter. Note that if $g$ is nonlinear there may be equality constraint violations during the variable space completion due to the use of numerical methods. A corresponding penalty for equality constraints can also be included.

\subsection{Physics-informed rounding (PhyR)} \label{sec:phyR}
The underlying concept of the Si\textbf{PhyR} framework is an explicit embedding of discrete decisions into the neural framework. Traditional optimization literature deals with MIPs using an array of heuristic methods to develop good upper and lower bounds, and explore and prune solution branches. We propose a physics-informed rounding algorithm taking inspiration from the class of rounding heuristics well-established in the MIP literature (\cite{MIP_50years}, and others including \cite{GoemansWilliamson_1995,Marchand_MIProunding_2001,Hifi_newrounding}). We consider the class of MIPs with binary variables~\eqref{eq:general_MIP} where the function $b$ uniquely defines a cutoff index $L$ such that $L$ variables have value 1, with the remaining $m-L$ variables having value 0. Our approach is as follows: function $b$ is used to define the cutoff index $L$ of a \textit{rounding function} $z_\tau = R_b(\hat{z}_\tau)$. This rounding function determines which $L$ variables are set to 1, and the remaining $m-L$ set to 0 based on the probabilities $\hat{z}_\tau$. To ensure gradient information is retained through \textbf{PhyR}, only $L-1$ variables will be rounded up, and $m-L-1$ variables rounded down. The neural training guides the remaining two variables to integer solutions. This is formally presented in Algorithm~\ref{alg:phyinfround}, and applied to power grid reconfiguration in Section~\ref{sec:phyR_forGrid}.

\RestyleAlgo{ruled}
\begin{algorithm}[H]
\DontPrintSemicolon
%  \State \textbf{Data: } $\hat{z}_\tau$
 \KwData{Probabilities $\hat{z}_\tau=\mathbb{P}(z_\tau = 1)$}
%  \State \textbf{Result: } Binary variables $z_\tau$ for switch-state prediction and topology selection
 \KwResult{Binary variables $z_\tau$}
 initialization: $R_b \Rightarrow L = f(b(\psi))$ \;
 Sort $\hat{z}_\tau$ in descending order; assign $\mathbb{I}_{\hat{z}_\tau}$ the sorted indices of $\hat{z}_\tau$ \;
 Assign $L-1$ binary variables to have value 1: $z_{\tau} \left[ \mathbb{I}_{\hat{z}_\tau}^{1:L-1} \right] =  \max{\left\{ \right. \hat{z}_\tau\left[ \mathbb{I}_{\hat{z}_\tau}^{1:L-1} \right], 1}\left.\right\} $ \;
 Assign $m-L-1$ binary variables to be 0: $z_\tau\left[ \mathbb{I}_{\hat{z}_\tau}^{L+1:m} \right] = \min{\left\{ \right. \hat{z}_\tau\left[ \mathbb{I}_{\hat{z}_\tau}^{L+1:m} \right], 0}\left.\right\} $
 \caption{Physics-Informed Rounding \textbf{PhyR} for binary variables}\label{alg:phyinfround}
\end{algorithm}

\subsection{Extensions to the proposed Si\textbf{PhyR} method}
% \paragraph{Integer variables:} 
\textbf{Integer variables:} Algorithm~\ref{alg:phyinfround} can be extended for integer variables $z_\tau \in \mathbb{Z}_\ell^m$ where the function $b$ defines a set of cutoff indices $\{L_1, L_2,...,L_{\ell-1}\}$. The largest $L_{\ell-1}$ probabilities map to the largest integer variable in the set $\mathbb{Z}_\ell$, the smallest $L_1$ probabilities map to the smallest integer variable, and so forth. These cutoff indices define $R_b$, which converts probabilities to integer values on the set $\mathbb{Z}_\ell$. 

% \paragraph{Binary variables in the dependent set:}
\textbf{Binary variables in the dependent set:} The Si\textbf{PhyR} framework in Fig.~\ref{fig:neural_framework} restricts binary variables to the independent set $z$ and uses $R_b$ to convert probabilities into binary decisions. This is motivated by the fact that MIPs are often formulated for sequential decision processes where the decision over integer variables precedes the decision of continuous variables: ex. the grid topology is selected before enforcing power flow constraints; a set of workers are selected before allocating tasks. The Si\textbf{PhyR} framework is flexible and can be extended to binary variables in the dependent set (i.e. $\varphi=[\varphi_\tau, \varphi_{\setminus\tau}$]) by employing the \textbf{PhyR}-based rounding as a post-processing step, similar to corrective approaches employed for continuous variables \cite{deepopf, deepopfplus, donti_DC3}. Note that this approach changes the interpretation of the variables: $\varphi_{\setminus\tau}$ calculated using the variable space completion must be interpreted as a probability, which is then passed through the rounding function $R_b$ to recover a binary decision. The use of this extension should be well motivated by application. It must also be highlighted that the Si\textbf{PhyR} framework optimizes decisions over both integer and continuous variables simultaneously, as these decisions are intricately linked through the constraints. Additional integer variables which impose \textit{soft} constraints can be included in either the independent or dependent variable sets. These variables may be driven towards integer solutions using approximations of rounding functions (see InSi in Section~\ref{sec:experiments}) or penalized in the loss function. 
%However, this breaks the logical flow of decision making

% \paragraph{Additional integer variables:}
% \textbf{Additional integer variables:} 

\section{Grid-Si\textbf{PhyR}: application to power systems for dynamic reconfiguration}
In this section we propose Grid-Si\textbf{PhyR} which applies the physics-informed ML framework Si\textbf{PhyR} to grid reconfiguration. This problem asks the following question: ``Given a distribution grid with a set of switches and distributed generation, what is the most efficient grid topology and resource dispatch (power setpoints) that satisfies all loads, subject to power physics, generator, and topology constraints?'' A naive method would conduct an exhaustive search of the space of switch status permutations and optimize the resource dispatch upon each topology to find the optimal switch on/off decisions (topology) and dispatch solution with minimal electrical line losses. Instead, we propose the use of Si\textbf{PhyR} towards learning to optimize the reconfiguration problem. The grid reconfiguration problem can be cast as a MIP as below.
\begin{align}
    \min_{\psi\in\mathcal{P}_x} & \; f_x(\psi) = \sum_{(i,j) \in \mathcal{T}_D} (P_{ij}^2 + P_{ji}^2 + Q_{ij}^2 + Q_{ji}^2)R_{ij} \label{eq:objectivefnc}\\
    & P_j^G - P_j^L = \sum_{k: (j,k)}P_{jk} - \sum_{i:(i,j)}P_{ij}, \;  Q_j^G - Q_j^L = \sum_{k: (j,k)}Q_{jk} - \sum_{i:(i,j)}Q_{ij}, \quad \forall j\in \mathcal{B} \label{eq:pbalance} \\
    & v_j - v_i = - 2(R_{ij}(P_{ij}-P_{ji}) + X_{ij}(Q_{ij}-Q_{ji})), \quad \forall (i,j)\in \mathcal{T}_\text{D}\setminus\mathcal{T}_\text{D}^{sw} \label{eq:Ohm0}\\
    & z_{ij} + z_{ji} = 1, \quad \forall (i,j)\in \mathcal{T}_D \setminus \mathcal{T}_D^{sw} \label{eq:zconstLine} \\
    & z_{ij} + z_{ji} = y_{ij}, \quad \forall (i,j)\in \mathcal{T}_D^{sw} \label{eq:zconstSwitch} \\
    & \sum_{(i,j)\in \mathcal{T}_{D}^{sw}} y_{ij} = (N-1) - (M-M_{sw}) \label{eq:radial1}
\end{align}
Eq.~\eqref{eq:objectivefnc}-\eqref{eq:zconstSwitch} describes a general distribution grid as a graph $\Gamma(\mathcal{B},\mathcal{T}_D)$, where $\mathcal{B}$ is the set of N nodes and $\mathcal{T}_D\coloneqq\{(i,j)\}$ is the set of M edges. The subset of lines with switches $\mathcal{T}_D^{sw} \subset \mathcal{T}_D$ contains $M_{sw}$ switches which can be turned on/off to change the grid topology. The distribution grid is connected to the transmission grid via the point of common coupling (PCC). The power variables are the real and reactive power loads $P^L_j, Q^L_j$ and generation $P^G_j,Q^G_j$ at every node $j$, the squared magnitude of nodal voltages $v_j$, and directed power flow through the distribution lines $P_{ij}, P_{ji}, Q_{ij}, Q_{ji}$. The line parameters of resistance and reactance are denoted as $R_{ij}$ and $X_{ij}$ respectively. The grid topology variables are the switch status $y_{ij} \forall (i,j) \in \mathcal{T}_{D}^{sw}\subset \mathcal{T}_{D}$ (where 1 is closed and 0 is open), and the selection of direction of power flow through each line $z_{ij}, z_{ji}$. These variables are binary variables.

The objective function~\eqref{eq:objectivefnc} approximates the electrical line losses which when minimized ensures efficient grid operations. Constraints \eqref{eq:pbalance}-\eqref{eq:Ohm0} describe power flow using the Linearized DistFlow model, commonly used for distribution grid modeling \cite{Baran_reconfig_1989}. Constraint \eqref{eq:Ohm0} is Ohm's law across all lines which do not have switches, and constraints \eqref{eq:pbalance} are lossless power balance at every node. Constraints \eqref{eq:zconstLine}-\eqref{eq:radial1} describe topological constraints and are composed of integer variables. Notably, constraint \eqref{eq:radial1} describes the radiality constraint such that the number of total branches is $N-1$, an operating requirement for US distribution grids. Additional inequality constraints describe the feasible set $\mathcal{P}_x$ of voltage and generator limits, grid connectivity constraint, and the conditional constraints for power flow across switches. The full MIQP model is provided in Appendix~\ref{app:network_MIP} for reference.

Next we provide the details of Grid-Si\textbf{PhyR} including the variable space decomposition into independent and dependent variables and design of the rounding function $R_b$ for topology selection.

\subsection{Variable space decomposition} \label{sec:varDecomposition_forGrid}
The reconfiguration problem includes two sets of binary variables: $y_{ij}$ pertaining to topology selection and $\left\{z_{ij},z_{ji}\right\}$ pertaining to power flow direction. The topology selection is a critical decision imposing hard constraints on the binary nature of switch status, while the power flow direction is a soft constraint. To accommodate both hard and soft constraints, the independent variables $z$ are decomposed into continuous and binary variables, denoted by superscripts C and B, respectively. We also classify $z_{\tau}$ and $z_{\setminus\tau}$ as the set of switch variables ($y_{ij}$) and the set of other variables, respectively, noting that $z_{\tau} \subset z^B$. The variable decomposition for grid reconfiguration is summarized in \eqref{eq:decomposition}, where the set of independent and dependent variables are $z=\left[z^B, z^C \right]$ and $\varphi=\left[\varphi^B, \varphi^C \right]$ respectively.
\begingroup
\allowdisplaybreaks
\begin{align}
    x &= \left[\left\{ P_j^L, Q_j^L \; \vert \; j\in\mathcal{B}\setminus \text{PCC} \right\}\right], \nonumber \\
    z^B &= \left[ z_{ji}, \left\{ y_{ij} \; \vert \; (i,j)\in \mathcal{T}_{D,1:M_{sw}-1}^{sw}\right\} \right],\nonumber \\
    z^C &= \left[ \left\{ v_j \; \vert \; j\in\mathcal{B}\setminus \text{PCC}\right\}, P_{ji}, P_{ij}, Q_{ji}, \left\{ Q_{ij} \; \vert \; (i,j) \in \mathcal{T}_{D}^{sw} \right\}, P_{j^\#}^L, Q_{j^\#}^L \right], \nonumber \\
    \varphi^B &= \left[ z_{ij}, \left\{ y_{ij} \; \vert \; (i,j)\in \mathcal{T}_{D,-1}^{sw}  \right\} \right], \quad \varphi^C = \left[ P_j^G, Q_j^G, \left\{ Q_{ij} \; \vert \; (i,j) \in \mathcal{T}_{D}\setminus \mathcal{T}_{D}^{sw} \right\} \right]
    \label{eq:decomposition}
\end{align}
\endgroup
% \begingroup
% \allowdisplaybreaks
% \begin{align}
%     x &= \left[\left\{ P_j^L, Q_j^L \; \vert \; j\in\mathcal{B}\setminus j^\# \right\}\right] \nonumber \\
%     z^B &= \left[ z_{ji},  y_{ij} \right], \quad z^C = \left[ \left\{ v_j \; \vert \; j\in\mathcal{B}\setminus j^\#\right\}, P_{ji}, P_{ij}, Q_{ji}, \left\{ Q_{ij} \; \vert \; (i,j) \in \mathcal{T}_{D}^{sw} \right\}, P_{j^\#}^L, Q_{j^\#}^L \right] \nonumber \\
%     \varphi^B &= \left[ z_{ij}\right], \quad \varphi^C = \left[ P_j^G, Q_j^G, \left\{ Q_{ij} \; \vert \; (i,j) \in \mathcal{T}_{D}\setminus \mathcal{T}_{D}^{sw} \right\} \right]
%     \label{eq:decomposition}
% \end{align}
% \endgroup
The equality constraints governing power flow are used to calculate $\varphi^C$ from the independent variables $z^C$. These correspond to $P_j^G$ and $Q_{ij}$ for non-switch lines from constraints~\eqref{eq:pbalance}, and $Q_j^G$ from constraint~\eqref{eq:Ohm0}. The equality constraints governing grid topology are used to calculate $\varphi^B$ from $z^B$. In particular, the corresponding dependent variables $z_{ij}$ are calculated from constraints \eqref{eq:zconstLine}-\eqref{eq:zconstSwitch}. 
% The treatment of binary variables $y_{ij}$ is presented in Section~\ref{sec:phyR_forGrid}.

\textit{Implication 1:} With the proposed Si\textbf{PhyR} framework the grid topology is selected, then the power physics are enforced upon the topology via the variable space completion. This sequential decision making improves prediction performance by mimicking the simpler OPF problem on a fixed topology, in every offline training iteration and online prediction. This is corroborated by the results in Section~\ref{sec:results}.

\textit{Implication 2:} A critical feature of the proposed variable decomposition is the certified satisfiability of inequality constraints describing voltage, line flow, and generator limits. By selecting voltages $v$ as an independent variable, these are scaled onto the box constraints describing operating limits. For a grid operator, this means voltage limits across the grid will always be satisfied, a critical aspect of power systems operation. This is inherent in our proposed structure, as compared to other methods which rely on projections, clipping, or penalties to enforce voltage constraints. 

\textit{Implication 3:} The equality constraints of the Linearized DistFlow model permit the dependent variables to be determined trivially with zero error. For problems which involve more complex (potentially nonlinear) equality constraints, such as nonlinear DistFlow or the full AC power flow, the same variable space reduction techniques can be used, and programs like Newton's method can be leveraged to solve for the dependent variables. In the backpropagation step through the equality constraint layer, the Jacobians describing the derivatives can be explicitly written out and the implicit function theorem used to backpropagate through the dependent variables. 

\subsection{Topology selection using \textbf{PhyR}} \label{sec:phyR_forGrid}
Distribution grids in the US are operated with a radial structure. Constraint~\eqref{eq:radial1} restricts the number of closed switches in the grid so it is radial with $N-1$ total branches, where $L$ switches must be closed and the remaining $M_{sw}-L$ must be open. This is a necessary condition for radiality. As it pertains to the \textbf{PhyR} method, the radiality constraint~\eqref{eq:radial1} is the function $b$ in Eq.~\eqref{eq:general_MIP}. The cutoff index of the rounding function $R_b$ is $L=(N-1) - (M-M_{sw})$. Embedding the above constraints in Grid-Si\textbf{PhyR} permits the selection of a feasible (i.e. radial) grid topology upon which the power flow describing the relationship between $z^C$ and $\varphi^C$ are satisfied (\textit{Implication 1} in the prior section).
% connected. <-- took this out since it's part of the inequality constraints

% Various mathematical formulations of these radiality and connectivity constraints include constraints on the determinant of the branch-to-node incidence matrix or spanning tree constraints and other graph theoretic approaches \cite{Lei_radiality_2020, Wang_radiality_2020, Ahmadi_radiality_2015, Lavorato_radiality_2012}. However, many of these suffer from high computational requirements and additional complexity, and do not leverage the fact that grid connectivity can be ensured by power flow constraints under normal operation (see Appendix~\ref{app:full_reconfig_MIP} for more details).
% \textcolor{red}{Constraint~\eqref{eq:radial2} enforces connectivity by requiring power to flow into or out of a node along at least one line. It should be noted that typical reconfiguration problem statements also include an arborescence constraint \cite{Taylor_reconfigModel}, either explicitly or implicitly in the formulation of the radiality constraint. However, the increasing penetration of DERs voids this assumption, and multiple generating sources (roots of the tree) must be permitted. We have relaxed this arborescence constraint in~\eqref{eq:radial2}.}

\section{Experiments} \label{sec:experiments}
We develop and implement multiple neural architectures to compare the performance of the proposed Grid-Si\textbf{PhyR}  (referred to by abbreviations in the result tables):
\begin{itemize}[leftmargin=0.6cm]
    \item \textbf{Optimizer:} Traditional optimization solver, Gurobi, a state-of-art commercial solver for MIPs. The optimal solution determined by the optimizer is denoted by the asterisk, i.e. $z^\ast,\varphi^\ast$.
    \item \textbf{InSi:} Without the proposed \textbf{PhyR} layer. Integer solutions for the switch status are encouraged (read: not enforced) by using a differentiable relaxation of the step function, the integer sigmoid (InSi): $\sigma_{InSi}(z) = \left[ 2\frac{1+\mu}{\mu+e^{-\tau z}} -1 \right]_{+}$, where $\tau, \mu$ are free parameters \cite{cao_sigmoidal}. InSi is used for the binary variables $\hat{z}^B$, while the traditional sigmoid is used for the continuous variables $\hat{z}^C$. The InSi function is parameterized by $\tau$ which governs the sharpness, i.e. how well it approximates the step function. Larger values of $\tau$ better discriminate between binary values, but render the function less differentiable and thus learning more challenging. We set the parameters $\mu=1$ and $\tau=5$ based on initial tests.
    \item \textbf{InSi2R:} Use the InSi architecture during training. During testing, the predictions $z_\tau$ are rounded to binary values before using the variable space completion.
    \item A set of \textbf{PhyR}-based methods with different output layer functions: Cla\textbf{PhyR} with a clamp, Si\textbf{PhyR} with a sigmoid, and InSi\textbf{PhyR} with the InSi function. We enforce the soft constraints on variables $z^B\setminus z_{\tau}$ using the $\sigma_{InSi}$ function.
\end{itemize}

We compare the performance of the end-to-end framework with supervised training, wherein the neural network has full knowledge of the optimal solution during training:
\begin{itemize}[leftmargin=0.6cm]
    \item \textbf{Supervised-x:} method `x' is trained with typical regression loss function: $l_{sup}(z,\varphi) = \Vert {(V_j - V_j^{\ast})^2 + (P_j^G - P_j^{G\ast})^2 + (Q_j^G - Q_j^{G\ast})^2}\Vert_2^2 + \Vert {(y_{ij}-y_{ij}^\ast)^2}\Vert_2^2$
    \item \textbf{Supervised-x-pen:} method `x' is trained with a soft loss penalty on inequality constraint violation: $l_{sup-pen}(z,\varphi) = l_{sup}(z,\varphi) + \lambda_h \Vert \text{max}\left\{0, h_x(z,\varphi)\right\} \Vert_2^2$
\end{itemize}

We also evaluated Grid-Si\textbf{PhyR} as a warm-start technique for traditional optimizers, but found minimal performance improvement (see Appendix~\ref{app:all_experiment_results}). Generally, the benefits seen for warm-start of general optimization are not easily replicated for MIPs.

\subsection{Grid reconfiguration datasets}
We consider two canonical distribution grids used in literature: \textit{BW-33} a synthetic grid with 33 nodes and 8 switches \cite{Baran_reconfig_1989}; and \textit{TPC-94} a model of a real distribution grid with 94 nodes and 14 switches \cite{Su_2005_83dataset}. These present excellent test cases for grid reconfiguration. The BW-33 is highly lossy with losses up to 8\% of total load and frequently violates grid voltage limits. The TPC-94 has 11 individual distribution feeders which can be connected to one another via switches (i.e. reconfiguration) to share load across feeders. We develop datasets by introducing distributed solar generation throughout the grid (to a penetration of 25\% generation-to-peak-load), and introduce load and generation profiles. We develop five different datasets for TPC-94: DS1 with loads perturbed about the nominal (typically done in literature \cite{vpoor_opf_topologyreconfig,Zamzam_SmartGridComm2020,Fioretto_lagrangeduals}); DS2 with residential load profiles; DS3 with residential and commercial load profiles; DS4 with has solar generation located in different nodes of the grid; and DS5 with residential load profiles but no solar generation profiles. Notably DS2 through DS5 are datasets with representative load and generation profiles. The datasets are presented in detail in Appendix~\ref{app:datasets}.

\subsection{Simulation parameters and hyperparameter configurations}
We use a lightweight neural architecture (to which Si\textbf{PhyR} is agnostic) with two fully connected hidden layers with linear transformation with bias, batch normalization, and ReLU activation. The size of input and output layers are determined by $x$ and $\hat{z}$. We backpropagate using ADAM with parameter $\gamma = 0.001$. The soft loss hyperparameter is set as $\lambda_h=100$, chosen to enforce a high penalty on violating inequality constraints while still allowing the underlying objective function to be improved upon. All datasets are split as $80\%$ training, and $10\%$ testing/validation. We use mini-batching with $200$ batch size. The learning rate ($lr=1e-3$ unless otherwise stated) and width of the neural network (5 for BW-33 and 300 for TPC-94) are determined through hyperparameter tuning. Training the InSi method and on representative datasets (DS-2 thru DS-5) requires a lower learning rate ($lr=1e-4$). Details of the hyperparameter tuning is provided in Appendix~\ref{app:hyperparam_tuning}. A committee machine approach is taken to evaluate the neural architectures, where 10 models are trained with independent weight initialization (all using He initialization \cite{He_initialization_2015}). An ensemble average is used by linearly combining the predictions across all 10 predictors. All neural architectures were developed and tested using PyTorch on an Apple M2 Max with 12-core CPU and 96GB RAM.

\subsection{Neural architecture performance metrics}
We assess our method on the following optimality (Opt) and feasibility (Feas) metrics:
% \textcolor{red}{Add explanation of the metrics}
\begin{itemize}[leftmargin=0.6cm]
    \item \textbf{[Opt] Dispatch error (DispErr):} mean-squared error (MSE) in optimal generator dispatch: 
    $\frac{1}{N}\sum_{j\in \mathcal{B}}{(P_j^G - P_j^{G\ast})^2 + (Q_j^G - Q_j^{G\ast})^2}$
    \item \textbf{[Opt] Voltage error (VoltErr):} MSE in nodal voltage prediction: $\frac{1}{N}\sum_{j\in \mathcal{B}}{(V_j - V_j^{\ast})^2}$
    \item \textbf{[Opt] Topology error (TopErr):} the Hamming distance \cite{Hamming_1986} between two topologies, calculated as the ratio of switch decisions not in the optimal position: 
    $\frac{1}{M_{sw}}\sum_{(i,j)\in \mathcal{T}_D^{sw}}{(y_{ij}-y_{ij}^\ast)^2}$.
    For \textbf{PhyR}-based methods the topology error is equivalent to using an indicator function which returns 1 if the switch status is not optimally selected, and 0 otherwise: 
    $\frac{1}{M_{sw}}\sum_{(i,j)\in \mathcal{T}_D^{sw}}{\mathbb{I}_{y_{ij}\neq y_{ij}^\ast}}$
    \item \textbf{[Feas] Inequality violation:} magnitude of violations in constraint set, measuring the mean and maximum as $\frac{1}{|h_x|}\sum_{k}{\max{\{ 0, h_x^k(\psi) \}}}$ and $\max_k{\{\max{\{ 0, h_x^k(\psi) \}}\}}$ respectively
    \item \textbf{[Feas] Number of violations exceeding a threshold:} the number of inequality constraints which are violated by more than an $\epsilon$ threshold: $\sum_{k}{\mathbb{I}_{\max{\{ 0, h_x^k(\psi) \}} > \epsilon}}$
\end{itemize}

\section{Results} \label{sec:results}
% \subsection{Performance of Grid-SiPhyR}
Table~\ref{tab:results_BW_33} compares the performance of Grid-Si\textbf{PhyR} with Optimizer and other learning approaches for the BW-33 grid. Table~\ref{tab:results_TPC_DS1} scales Grid-Si\textbf{PhyR} to the larger TPC-94 grid and assesses performance on different datasets. Smaller values for each metric are better and indicates performance closer to Optimizer and thus closer to both optimality and feasibility. 

% \paragraph{Grid-SiPhyR preserves feasibility and achieves reasonable performance on optimality:}
\textbf{Grid-SiPhyR preserves feasibility and achieves reasonable performance on optimality:} We note that from Table~\ref{tab:results_BW_33}, Grid-Si\textbf{PhyR} consistently outperforms other methods by better preserving feasibility with respect to inequality constraints (5 inequality violations versus 14, with a ten-fold reduction in both mean and maximum violation magnitude), while achieving reasonable generator dispatch and nodal voltage predictions compared to optimal. These results follow from the fact that the neural predictor in our Si\textbf{PhyR} approach uses the \textbf{PhyR} layer to select a feasible topology for each training instance upon which it predicts a feasible power flow solution (i.e. equality constraints are satisfied) and learns to reduce inequality violations while improving upon optimality objectives. Without \textbf{PhyR} in the training loop (as with InSi and InSi2R) the neural predictor is unable to preserve feasibility of inequality constraints: rather than considering a single topology, it must satisfy power flow across a (potentially large) set of possible topologies (i.e. $z_\tau$ takes on non-binary values when \textbf{PhyR} is not used). Not only is this a challenging task, but there may not exist a power flow solution which remains feasible across multiple topologies. Even with explicit selection of a topology during testing (but not training) as with InSi2R, neither feasibility nor optimality can be improved upon, as clear by the same DispErr and VoltErr for InSi and InSi2R (rows 1 and 2 of Table~\ref{tab:results_BW_33}). A comment must be made on the TopErr. Model variance is a challenge in reconfiguration prediction, with significant spread in TopErr across multiple predictors. Prediction performance can be improved substantially when considering the best performing predictor in the ensemble (as shown in brackets): the TopErr for Si\textbf{PhyR} reduces from 41.5\% to 13.7\%, and now has comparable performance to the Supervised-InSi method which has full access to the optimal solution. Grid-Si\textbf{PhyR} then produces the lowest TopErr of all tested methods. 

% \paragraph{Grid-SiPhyR can learn to optimize while preserving feasibility:}
\textbf{Grid-SiPhyR can learn to optimize while preserving feasibility:} Table~\ref{tab:results_BW_33} shows that the proposed end-to-end learning framework can obtain comparable performance to a supervised approach. This ability of Grid-Si\textbf{PhyR} to \textit{learn to optimize} is essential to applications in power systems, where the optimal solution is typically unknown and is computationally prohibitive to obtain for training data. Clearly, supervised methods have lower optimality errors as they are given full knowledge of the optimal solution. At this point it is necessary to distinguish between optimality and feasibility. The power system is a critical infrastructure and meeting electricity demand with available generation is a critical action. Thus feasibility is vastly more important than optimality: only when we have feasibility, can we begin to think about optimality. This indisputable prioritization gives a clear criterion for evaluating the performance across different methods. Physics-informed approaches like Grid-Si\textbf{PhyR} which prioritizes feasibility explicitly in the design result in the lowest feasibility errors (even compared to Supervised-x-pen) and has higher prediction accuracy. % (decidedly in dispatch and voltage, similar in topology)

% \paragraph{Grid-SiPhyR can scale to larger grids and generalize across load datasets:}
\textbf{Grid-SiPhyR can scale to larger grids and generalize across datasets:} Table~\ref{tab:results_TPC_all} shows that once again the proposed \textbf{PhyR}-based predictors outperforms InSi2R on all metrics (particularly for feasibility), especially when tested on different datasets than training. It is also immediately obvious that the prediction performance is significantly enhanced by training on representative load and generation datasets (i.e. DS2 and DS5). Comparing the error metrics for DS1 on DS3 (top right) with DS2 on DS3 (middle right), the errors are significantly reduced for the latter: 40\% lower DispErr, 95\% lower VoltErr, 87\% lower Mean ineq, and the number of inequality violations reduces from 8\% to 4\%. Since real-world data of network topologies and customer load profiles are not widely available due to security and privacy concerns, the creation of synthetic datasets with representative load profiles and DER penetration requires considerable effort, but is a valuable and necessary task. The datasets generated in this work are a key contribution towards supporting the development, testing, and comparison of algorithms for power systems control and optimization.

% We find that DC3 preserves feasibility with respect to both equality and inequality constraints, while
% achieving reasonable objective values. (The average per-instance optimality gap for DC3 over the
% classical optimizer is 10.59%.) For every baseline deep learning algorithm, on the other hand,
% feasibility is violated significantly for either equality or inequality constraints. As expected, “DC3
% 6=” (completion ablated) results in violated equality constraints, while “DC3 6≤” (correction ablated)
% violates inequality constraints. Ablating the soft loss also results in violated inequality constraints,
% leading to an objective value significantly lower than would be possible were constraints satisfied.

% \textcolor{red}{Highlight some of the performance elements in blue or red to indicate good and bad performance}

\begin{table}[]
\centering 
% \footnotesize
% \scriptsize
\caption{Results for the BW-33 grid tested on 876 instances. The committee machine performance (i.e. ensemble average over 10 trained models) on each metric averaged over all test instances is shown. The performance of the predictor with the lowest TopErr is shown in brackets for select metrics. Lower values are better for all metrics.}
\label{tab:results_BW_33}
\begin{tabular}{@{}lccccccc@{}}
\toprule
\multicolumn{1}{c}{\multirow{2}{*}{Method}} & \multicolumn{6}{c}{Metric} \\ \cmidrule(l){2-7} 
\multicolumn{1}{c}{} & DispErr & VoltErr & TopErr & Mean ineq & Max ineq & Num ineq \\ \cmidrule(r){1-1}
InSi        & 3.24\text{e-}2 (3.44\text{e-}2) & 2.30\text{e-}3 & 49.7\% (23.2\%) & 1.53\text{e-}3 & \textcolor{red}{0.148} & \textcolor{red}{16.3 (18.3)} \\
InSi2R      & 3.24\text{e-}2 (3.44\text{e-}2) & 2.30\text{e-}3 & 48.6\% (19.9\%) & 1.02\text{e-}3 & \textcolor{red}{0.114} & \textcolor{red}{14.1 (15.3)} \\
Cla\textbf{PhyR}     & 2.86\text{e-}2 (3.29\text{e-}2) & 1.12\text{e-}3 & 52.4\%(42.1\%) & 3.80\text{e-}4 & 2.40\text{e-}2 & 4.42 (2.94) \\
Si\textbf{PhyR}      & \textcolor{blue}{2.89\text{e-}2 (8.92\text{e-}3)} & \textcolor{blue}{1.69\text{e-}3} & 41.5\%\textcolor{blue}{(13.7\%)} & \textcolor{blue}{4.79\text{e-}4} & \textcolor{blue}{4.23\text{e-}2} & \textcolor{blue}{5.72 (3.94)} \\
InSi\textbf{PhyR}    & 3.18\text{e-}2 (3.72\text{e-}2) & 1.13\text{e-}3 & 44.6\%(21.6\%) & 5.00\text{e-}4 & 4.88\text{e-}2 & 6.02 (5.82) \\
\hline 
Supervised-InSi  & 8.61\text{e-}4 & 4.65\text{e-}4 & \textcolor{blue}{14.2\%} & 5.52\text{e-}3 & \textcolor{red}{0.817} & \textcolor{red}{37.4} \\
Supervised-\textbf{PhyR}  & 1.51\text{e-}4 & 2.36\text{e-}4 & 29.9\% & 4.29\text{e-}3 & \textcolor{red}{0.888} & \textcolor{red}{19.6} \\
Supervised-InSi-pen & 1.00\text{e-}3 & 2.80\text{e-}3 & 25.8\% & 1.55\text{e-}3 & \textcolor{red}{0.173} & \textcolor{red}{15.7} \\
Supervised-\textbf{PhyR}-pen & 5.78\text{e-}4 & 1.35\text{e-}3 & 33.6\% & \textcolor{blue}{5.49\text{e-}4} & \textcolor{blue}{4.84\text{e-}2} & \textcolor{blue}{7.26} \\
 \bottomrule
\end{tabular}
\end{table}

\begin{table} [h!]
\centering
% \scriptsize
\caption{Results for the TPC-94 grid tested on 8640 instances. All networks were trained on the DS1 (perturbed loads), DS2 (residential loads), and DS5 (residential loads without solar profiles) datasets, and are \textbf{tested} on datasets DS2 and DS3 (mixed residential and commercial loads). Lower values are better for all metrics.}
\label{tab:results_TPC_all}
\begin{tabular}{lcccccccc}
\toprule
\multicolumn{1}{l}{\multirow{2}{*}{\begin{tabular}[c]{@{}c@{}}Method\\ (trained on DS1\end{tabular}}}
& \multicolumn{4}{c}{DS2, residential loads}   & \multicolumn{4}{c}{DS3, mixed loads}   \\
\cmidrule(lr){2-5} \cmidrule(lr){6-9}
&
\multicolumn{1}{c}{DispErr} & \multicolumn{1}{c}{VoltErr}     & \multicolumn{1}{c}{Mean ineq} & \multicolumn{1}{c}{Num ineq }     &
\multicolumn{1}{c}{DispErr} & \multicolumn{1}{c}{VoltErr}     & \multicolumn{1}{c}{Mean ineq} & \multicolumn{1}{c}{Num ineq } \\
\midrule
InSi2R      & \textcolor{red}{1.03} & \textcolor{red}{1.47\text{e-}2} & \textcolor{red}{3.20\text{e-}2} & \textcolor{red}{165} & \textcolor{red}{1.10} & \textcolor{red}{1.02\text{e-}2} & \textcolor{red}{3.69\text{e-}2} & \textcolor{red}{166} \\
Cla\textbf{PhyR}     & 1.77\text{e-}2 & 3.55\text{e-}3 & 4.43\text{e-}3 & 107 & 1.99\text{e-}2 & 3.41\text{e-}3 & 5.01\text{e-}3 & 122 \\
Si\textbf{PhyR}      & \textcolor{blue}{1.72\text{e-}2} & \textcolor{blue}{2.91\text{e-}3} & \textcolor{blue}{4.33\text{e-}3} & \textcolor{blue}{106} & \textcolor{blue}{1.67\text{e-}2} & \textcolor{blue}{2.16\text{e-}3} & \textcolor{blue}{4.47\text{e-}3} & \textcolor{blue}{118}\\
\midrule
\multicolumn{1}{l}{\multirow{2}{*}{\begin{tabular}[c]{@{}c@{}}Method\\ (trained on DS2\end{tabular}}} & \multicolumn{4}{c}{DS2, residential loads}   & \multicolumn{4}{c}{DS3, mixed loads}   \\
\cmidrule(lr){2-5} \cmidrule(lr){6-9}
&
\multicolumn{1}{c}{DispErr} & \multicolumn{1}{c}{VoltErr}     & \multicolumn{1}{c}{Mean ineq} & \multicolumn{1}{c}{Num ineq }     &
\multicolumn{1}{c}{DispErr} & \multicolumn{1}{c}{VoltErr}     & \multicolumn{1}{c}{Mean ineq} & \multicolumn{1}{c}{Num ineq } \\
\midrule
InSi2R      & 1.12\text{e-}2 & 3.98\text{e-}3 & 1.94\text{e-}3 & 71.2 & 2.98\text{e-}2 & 2.22\text{e-}3 & 5.36\text{e-}3 & 108\\
Cla\textbf{PhyR}     & \textcolor{blue}{1.00\text{e-}2} & \textcolor{blue}{2.38\text{e-}3} & \textcolor{blue}{5.07\text{e-}4} & \textcolor{blue}{21.6} & \textcolor{blue}{9.987\text{e-}3} & \textcolor{blue}{1.91\text{e-}3} & \textcolor{blue}{1.40\text{e-}3} & \textcolor{blue}{64.7} \\
\midrule
\multicolumn{1}{l}{\multirow{2}{*}{\begin{tabular}[c]{@{}c@{}}Method\\ (trained on DS5\end{tabular}}} & \multicolumn{4}{c}{DS2, residential loads}   & \multicolumn{4}{c}{DS3, mixed loads}   \\
\cmidrule(lr){2-5} \cmidrule(lr){6-9}
&
\multicolumn{1}{c}{DispErr} & \multicolumn{1}{c}{VoltErr}     & \multicolumn{1}{c}{Mean ineq} & \multicolumn{1}{c}{Num ineq }     &
\multicolumn{1}{c}{DispErr} & \multicolumn{1}{c}{VoltErr}     & \multicolumn{1}{c}{Mean ineq} & \multicolumn{1}{c}{Num ineq } \\
\midrule
InSi2R      & 1.05\text{e-}2 & 3.86\text{e-}3 & 1.45\text{e-}3 & 64.3 & 1.29\text{e-}2 & 2.22\text{e-}3 & 3.04\text{e-}3 & 104 \\
Cla\textbf{PhyR}     & \textcolor{blue}{9.18\text{e-}3} & \textcolor{blue}{2.71\text{e-}3} & \textcolor{blue}{8.92\text{e-}4} & \textcolor{blue}{39.1} & \textcolor{blue}{9.94\text{e-}3} & \textcolor{blue}{2.08\text{e-}3} & \textcolor{blue}{1.78\text{e-}3} & \textcolor{blue}{79.2} \\
\bottomrule
\end{tabular}
\end{table}

\textbf{Grid-SiPhyR is an enabling technology for essential grid operations in future decarbonized energy systems:} Simulation results on the two canonical distribution grids show that the dynamic reconfiguration paradigm enabled by Grid-Si\textbf{PhyR} contributes to three essential power system design goals: improves grid efficiency by reducing electrical losses, ensures grid operability by improving voltage profiles, and supports clean energy directives by increasing solar energy utilization. Specifically, dynamic reconfiguration can reduce line losses by 23\% (320 MWh, powers an additional 30 US households for a year). It can also significantly improve voltage profiles in heavily loaded networks by increasing network-wide voltages and reducing the number of voltage violations by as much as 66\%. Finally, the use of dynamic reconfiguration allows local solar generation to be directly connected to loads which increases solar utilization by 6\% (250 MWh, powers 23 US households with clean energy for a year). This increase in local solar utilization reduces reliance on electricity imported from the bulk system. In regions where fossil-based generation dominates the energy mix, this corresponds to reductions in CO\textsubscript{2} emissions; for Massachusetts (state in the US), the reduction is about 107 metric tons of CO\textsubscript{2} per year. Detailed simulation results are presented in Appendix~\ref{app:powersystems_results}.

% EIA report -- Massachusetts emissions from electricity generation in 2021 is 974 lbs/MWh. See more here: https://www.eia.gov/electricity/state/massachusetts/

\section{Conclusion}
We have presented a general framework, Si\textbf{PhyR} for end-to-end learning to optimize for combinatorial problems. Our approach integrates a rounding function directly within a physics-informed differentiable framework to ensure certified satisfiability of critical equality constraints and binary variables. We apply our method to the power grid and show that Grid-Si\textbf{PhyR} can learn to optimize and provide good and fast solutions while preserving feasibility of key constraints. Future work will consider formally extending the Si\textbf{PhyR} framework for integer variables and set constraints. We will also explore the development of Grid-Si\textbf{PhyR} to generalize to new grid topologies using GNNs or transfer learning techniques. Finally, we will explore the use of generative adversarial networks for developing diverse and representative datasets for power systems optimization and control.

\newpage 
\bibliography{references}

\newpage
\appendix

\section{Additional experiment results and warm-start results} \label{app:all_experiment_results}
Extensive evaluation of the proposed physics-informed framework were carried out with detailed discussion presented in Section~\ref{sec:results} of the main paper. Additional simulation results are presented in this Appendix. 

\subsection{Summary of key results}
The key results presented in Section~\ref{sec:results} are repeated here for completion.
\begin{enumerate}[start=1,label={\upshape\bfseries Result \arabic*:},wide = 0pt, leftmargin = 4.2em]
% [start=1,label={\bfseries Result \arabic*:}]
    \item Physics-informed methods enable higher prediction accuracy in both optimality and feasibility metrics; %Phy-R based methods outperform InSi in both optimality and feasibility metrics
    % \item There is a clear trade-off between optimality versus feasibility;
    \item The proposed Si\textbf{PhyR} method can \textit{learn to optimize}. Supervised learning outperforms unsupervised learning in optimality metrics, but underperforms in feasibility metrics; % reframe to say something like "unsupervised methods can be suitable..." ?
    \item Datasets used for algorithm development must be representative of real system load and generation characteristics;
    \item Model variance is a challenge in reconfiguration prediction, with significant spread in prediction accuracy of topology error across multiple predictors
    % \item Smoothness of the output layer function impacts training and prediction performance, with SiPhyR outperforming other architectures
    \item Warm-start of MIPs is challenging: warm-start with the neural prediction can improve worst-case performance, but additional tuning is needed to further reduce solve time. Even then, there are no guarantees on optimizer performance.
\end{enumerate}

Recall that for all experiments a committee machine approach was taken where 10 models were trained with independent weight initialization (following He initialization) for each parameter. Unless otherwise indicated, the results presented are the average over the 10 predictors.

Table~\ref{tab:results_BW_33_maxpredictor} presents the performance results for the predictor with the best performance on topology error for the BW-33 grid. This predictor shows significant improvement in prediction accuracy across all metrics, and in particular for the topology error (TopErr), as compared to the ensemble average over all committee members (see Table~\ref{tab:results_BW_33}).

Table~\ref{tab:results_TPC_DS1} thru Table~\ref{tab:results_TPC_DS5} present results for the TPC-94 grid. The predictors are trained on datasets DS1, DS2, or DS5, and tested on datasets DS1, DS2, DS3, and DS4. The results showcase the scalability and superior performance of the proposed Si\textbf{PhyR} method, and indicate the importance of representative training data, as discussed in Section~\ref{sec:results}.

\begin{table}[h]
\centering
\caption{Results for the BW-33 grid tested on 876 instances. The performance for the predictor with the lowest TopErr is presented (i.e. a single predictor which performs the best among the 10 predictors trained in the committee). Lower values are better for all metrics.}
\label{tab:results_BW_33_maxpredictor}
\begin{tabular}{@{}lccccccc@{}}
\toprule
\multicolumn{1}{c}{\multirow{2}{*}{Method}} & \multicolumn{7}{c}{Metric} \\ \cmidrule(l){2-8} 
\multicolumn{1}{c}{} & DispErr & VoltErr & TopErr & Max TopErr & Mean ineq & Max ineq & Num ineq \\ \cmidrule(r){1-1}
InSi   & 3.44\text{e-}2 & 4.01\text{e-}3 & 23.2\% & 100\% & 2.22\text{e-}3 & 0.190 & 18.3\\
InSi2R  & 3.44\text{e-}2 & 4.01\text{e-}3 & 19.9\% & 79.5\% & 1.34\text{e-}3 & 0.147 & 15.3\\
Cla\textbf{PhyR}   & 3.29\text{e-}2 & 1.21\text{e-}3 & 42.1\% & 75.0\% & 3.06\text{e-}4 & 2.06\text{e-}2 & 2.94\\
Si\textbf{PhyR}   & 8.92\text{e-}3 & 3.06\text{e-}3 & 13.7\% & 25.0\% & 3.56\text{e-}4 & 2.56\text{e-}2 & 3.94\\
InSi\textbf{PhyR}   & 3.72\text{e-}2 & 6.68\text{e-}4 & 21.6\% & 50.0\% & 5.06\text{e-}4 & 5.67\text{e-}4 & 5.82\\
\bottomrule
\end{tabular}
\end{table}

\begin{table}[]
\centering 
\caption{Results for the TPC-94 grid tested on 8760 instances. All networks were trained on the DS1 (perturbed loads) dataset and are \textbf{tested} on datasets DS1, DS2, DS3, and DS4. Lower values are better for all metrics.}
\label{tab:results_TPC_DS1}
\begin{tabular}{@{}lcccccc@{}}
\toprule
\multicolumn{1}{c}{\multirow{2}{*}{\begin{tabular}[c]{@{}c@{}}Method\\ (tested on DS1, perturbed)\end{tabular}}} & \multicolumn{6}{c}{Metric} \\ \cmidrule(l){2-7} 
\multicolumn{1}{c}{} & DispErr & VoltErr & TopErr & Mean ineq & Max ineq & Num ineq \\ \cmidrule(r){1-1}
InSi        & 1.41 & 3.31\text{e-}2 & 44.3\% & 3.21\text{e-}2 & 3.01 & 162\\
InSi2R      & 1.41 & 3.31\text{e-}2 & 43.6\% & 3.20\text{e-}2 & 2.98 & 161 \\
PhyR        & 1.78\text{e-}2 & 1.52\text{e-}2 & 45.4\% & 8.11\text{e-}4 & 5.25\text{e-}2 & 40.5\\
Cla\textbf{PhyR}     & 1.15\text{e-}2 & 1.31\text{e-}2 & 44.1\% & 7.67\text{e-}4 & 4.63\text{e-}2 & 41.1\\
Si\textbf{PhyR}      & 1.12\text{e-}2 & 1.51\text{e-}2 & 45.4\% & 7.27\text{e-}4 & 4.25\text{e-}2 & 37.5\\
InSi\textbf{PhyR}    & 1.10\text{e-}2 & 1.39\text{e-}2 & 44.4\% & 7.87\text{e-}4 & 4.62\text{e-}2 & 42.5\\
 \toprule
 \multicolumn{1}{c}{\multirow{2}{*}{\begin{tabular}[c]{@{}c@{}}Method\\ (tested on DS2, residential)\end{tabular}}} & \multicolumn{6}{c}{Metric} \\ \cmidrule(l){2-7} 
\multicolumn{1}{c}{} & DispErr & VoltErr & TopErr & Mean ineq & Max ineq & Num ineq \\ \cmidrule(r){1-1}
InSi        & 1.03 & 1.47\text{e-}2 & 41.6\% & 3.22\text{e-}2 & 2.61 & 166\\
InSi2R      & 1.03 & 1.47\text{e-}2 & 40.9\% & 3.20\text{e-}2 & 2.61 & 165 \\
PhyR        & 1.71\text{e-}2 & 4.80\text{e-}3 & 43.7\% & 4.32\text{e-}3 & 0.960 & 107\\
Cla\textbf{PhyR}     & 1.77\text{e-}2 & 3.55\text{e-}3 & 43.3\% & 4.43\text{e-}3 & 1.00 & 107 \\
Si\textbf{PhyR}      & 1.72\text{e-}2 & 2.91\text{e-}3 & 46.6\% & 4.33\text{e-}3 & 0.959 & 106 \\
InSi\textbf{PhyR}    & 1.71\text{e-}2 & 5.18\text{e-}3 & 44.0\% & 4.25\text{e-}3 & 0.963 & 103 \\
 \toprule
 \multicolumn{1}{c}{\multirow{2}{*}{\begin{tabular}[c]{@{}c@{}}Method\\ (tested on DS3, mixed)\end{tabular}}} & \multicolumn{6}{c}{Metric} \\ \cmidrule(l){2-7} 
\multicolumn{1}{c}{} & DispErr & VoltErr & TopErr & Mean ineq & Max ineq & Num ineq \\ \cmidrule(r){1-1}
InSi        & 1.10 & 1.02\text{e-}2 & 42.8\% & 3.71\text{e-}2 & 3.09 & 167 \\
InSi2R      & 1.10 & 1.02\text{e-}2 & 41.6\% & 3.69\text{e-}2 & 3.09 & 166 \\
PhyR        & 1.63\text{e-}2 & 3.61\text{e-}2 & 45.0\% & 4.35\text{e-}3 & 0.960 & 117 \\
Cla\textbf{PhyR}     & 1.99\text{e-}2 & 3.41\text{e-}3 & 44.4\% & 5.01\text{e-}3 & 0.988 & 122\\
Si\textbf{PhyR}      & 1.67\text{e-}2 & 2.16\text{e-}3 & 44.7\% & 4.47\text{e-}3 & 0.958 & 118 \\
InSi\textbf{PhyR}    & 1.64\text{e-}2 & 3.87\text{e-}3 & 42.4\% & 4.31\text{e-}3 & 0.962 & 114 \\
 \toprule
 \multicolumn{1}{c}{\multirow{2}{*}{\begin{tabular}[c]{@{}c@{}}Method\\ (tested on DS4, solar error)\end{tabular}}} & \multicolumn{6}{c}{Metric} \\ \cmidrule(l){2-7} 
\multicolumn{1}{c}{} & DispErr & VoltErr & TopErr & Mean ineq & Max ineq & Num ineq \\ \cmidrule(r){1-1}
InSi        & 1.03 & 1.31\text{e-}2 & 41.2\% & 3.22\text{e-}2 & 2.62 & 166 \\
InSi2R      & 1.03 & 1.31\text{e-}2 & 40.5\% & 3.21\text{e-}2 & 2.62 & 165 \\
PhyR        & 1.71\text{e-}2 & 4.06\text{e-}3 & 43.7\% & 4.33\text{e-}3 & 0.960 & 107 \\
Cla\textbf{PhyR}     & 1.78\text{e-}2 & 3.01\text{e-}3 & 43.3\% & 4.42\text{e-}3 & 0.959 & 107 \\
Si\textbf{PhyR}      & 1.73\text{e-}2 & 2.38\text{e-}3 & 46.5\% & 4.33\text{e-}3 & 0.959 & 106 \\
InSi\textbf{PhyR}    & 1.71\text{e-}2 & 4.45\text{e-}3 & 44.0\% & 4.25\text{e-}3 & 0.963 & 103 \\
\bottomrule
\end{tabular}
\end{table}

\begin{table}[]
\centering
\caption{Results for the TPC-94 grid tested on 8760 instances. All networks were trained on the DS2 (residential loads) dataset and are \textbf{tested} on datasets DS1, DS2, DS3, and DS4. Lower values are better for all metrics.}
\label{tab:results_TPC_DS2}
\begin{tabular}{@{}lcccccc@{}}
\toprule
\multicolumn{1}{c}{\multirow{2}{*}{\begin{tabular}[c]{@{}c@{}}Method\\ (tested on DS2, residential)\end{tabular}}} & \multicolumn{6}{c}{Metric} \\ \cmidrule(l){2-7} 
\multicolumn{1}{c}{} & DispErr & VoltErr & TopErr & Mean ineq & Max ineq & Num ineq \\ \cmidrule(r){1-1}
InSi        & 1.12\text{e-}2 & 3.98\text{e-}3 & 47.8\% & 2.29\text{e-}3 & 0.324 & 73.1 \\
InSi2R      & 1.12\text{e-}2 & 3.98\text{e-}3 & 45.7\% & 1.94\text{e-}3 & 0.148 & 71.2 \\
Cla\textbf{PhyR}        & 1.00\text{e-}2 & 2.38\text{e-}3 & 48.6\% & 5.07\text{e-}4 & 5.57\text{e-}3 & 21.6 \\
 \toprule
\multicolumn{1}{c}{\multirow{2}{*}{\begin{tabular}[c]{@{}c@{}}Method\\ (tested on DS1, perturbed)\end{tabular}}} & \multicolumn{6}{c}{Metric} \\ \cmidrule(l){2-7} 
\multicolumn{1}{c}{} & DispErr & VoltErr & TopErr & Mean ineq & Max ineq & Num ineq \\ \cmidrule(r){1-1}
InSi        & 0.449 & 1.80\text{e-}2 & 47.5\% & 3.80\text{e-}2 & 2.95 & 192 \\
InSi2R      & 0.449 & 1.80\text{e-}2 & 45.5\% & 3.77\text{e-}2 & 2.95 & 190 \\
Cla\textbf{PhyR}        & 0.148 & 2.35\text{e-}2 & 46.2\% & 2.35\text{e-}2 & 2.36 & 168 \\
 \toprule
\multicolumn{1}{c}{\multirow{2}{*}{\begin{tabular}[c]{@{}c@{}}Method\\ (tested on DS3, mixed)\end{tabular}}} & \multicolumn{6}{c}{Metric} \\ \cmidrule(l){2-7} 
\multicolumn{1}{c}{} & DispErr & VoltErr & TopErr & Mean ineq & Max ineq & Num ineq \\ \cmidrule(r){1-1}
InSi        & 2.98\text{e-}2 & 2.22\text{e-}3 & 46.6\% & 5.58\text{e-}3 & 0.488 & 110 \\
InSi2R      & 2.98\text{e-}2 & 2.22\text{e-}3 & 44.3\% & 5.36\text{e-}3 & 0.358 & 108 \\
Cla\textbf{PhyR}        & 9.987\text{e-}3 & 1.91\text{e-}3 & 44.0\% & 1.40\text{e-}3 & 0.125 & 64.7 \\
 \toprule
\multicolumn{1}{c}{\multirow{2}{*}{\begin{tabular}[c]{@{}c@{}}Method\\ (tested on DS4, solar error)\end{tabular}}} & \multicolumn{6}{c}{Metric} \\ \cmidrule(l){2-7} 
\multicolumn{1}{c}{} & DispErr & VoltErr & TopErr & Mean ineq & Max ineq & Num ineq \\ \cmidrule(r){1-1}
InSi        & 1.14\text{e-}2 & 3.27\text{e-}3 & 47.5\% & 2.37\text{e-}3 & 0.327 & 74.0 \\
InSi2R      & 1.14\text{e-}2 & 3.27\text{e-}3 & 45.4\% & 2.02\text{e-}3 & 0.153 & 72.0 \\
Cla\textbf{PhyR}        & 1.00\text{e-}2 & 1.96\text{e-}3 & 48.2\% & 5.09\text{e-}4 & 5.55\text{e-}2 & 21.9 \\
\bottomrule
\end{tabular}
\end{table}

\begin{table}[]
\centering
\caption{Results for the TPC-94 grid tested on 8760 instances. All networks were trained on the DS5 (flat solar) dataset and are \textbf{tested} on datasets DS2, DS3, and DS4. Lower values are better for all metrics.}
\label{tab:results_TPC_DS5}
\begin{tabular}{@{}lcccccc@{}}
\toprule
\multicolumn{1}{c}{\multirow{2}{*}{\begin{tabular}[c]{@{}c@{}}Method\\ (tested on DS2, residential)\end{tabular}}} & \multicolumn{6}{c}{Metric} \\ \cmidrule(l){2-7} 
\multicolumn{1}{c}{} & DispErr & VoltErr & TopErr & Mean ineq & Max ineq & Num ineq \\ \cmidrule(r){1-1}
InSi        & 1.05\text{e-}2 & 3.86\text{e-}3 & 47.8\% & 1.80\text{e-}3 & 0.314 & 66.2 \\
InSi2R      & 1.05\text{e-}2 & 3.86\text{e-}3 & 44.8\% & 1.45\text{e-}3 & 0.120 & 64.3 \\
Cla\textbf{PhyR}        & 9.18\text{e-}3 & 2.71\text{e-}3 & 47.9\% & 8.92\text{e-}4 & 9.26\text{e-}2 & 39.1 \\
 \toprule
\multicolumn{1}{c}{\multirow{2}{*}{\begin{tabular}[c]{@{}c@{}}Method\\ (tested on DS3, mixed)\end{tabular}}} & \multicolumn{6}{c}{Metric} \\ \cmidrule(l){2-7} 
\multicolumn{1}{c}{} & DispErr & VoltErr & TopErr & Mean ineq & Max ineq & Num ineq \\ \cmidrule(r){1-1}
InSi        & 1.29\text{e-}2 & 2.22\text{e-}3 & 46.5\% & 3.33\text{e-}3 & 0.340 & 105 \\
InSi2R      & 1.29\text{e-}2 & 2.22\text{e-}3 & 44.3\% & 3.04\text{e-}3 & 0.230 & 104 \\
Cla\textbf{PhyR}        & 9.94\text{e-}3 & 2.08\text{e-}3 & 46.3\% & 1.78\text{e-}3 & 0.133 & 79.2 \\
 \toprule
\multicolumn{1}{c}{\multirow{2}{*}{\begin{tabular}[c]{@{}c@{}}Method\\ (tested on DS4, solar error)\end{tabular}}} & \multicolumn{6}{c}{Metric} \\ \cmidrule(l){2-7} 
\multicolumn{1}{c}{} & DispErr & VoltErr & TopErr & Mean ineq & Max ineq & Num ineq \\ \cmidrule(r){1-1}
InSi        & 1.05\text{e-}2 & 3.10\text{e-}3 & 47.8\% & 1.82\text{e-}3 & 0.314 & 67.1 \\
InSi2R      & 1.05\text{e-}2 & 3.10\text{e-}3 & 44.8\% & 1.47\text{e-}3 & 0.121 & 65.2 \\
Cla\textbf{PhyR}        & 9.16\text{e-}3 & 2.18\text{e-}3 & 47.9\% & 9.07\text{e-}4 & 9.37\text{e-}2 & 39.7 \\
\bottomrule
\end{tabular}
\end{table}

\subsection{Performance of Grid-Si\textbf{PhyR} as a warm-start technique}
Prior work has looked at developing machine learning models to predict warm-start points for traditional optimization solvers. This section investigates the ability for Si\textbf{PhyR} to act as a warm-start predictor. Before presenting the warm-start experiment details and result, it is necessary to note that warm-start points are \textit{generally} effective techniques to reduce optimization time. However, the benefits seen for general linear and nonlinear optimization are not replicated for mixed integer optimization problems. The techniques employed to solve the class of mixed integer problems, of which grid reconfiguration is a member, rely on multiple heuristics such as branch and bound, cutting planes, node presolve, and symmetry detection (among others). The selection of which techniques are used, the order in which they are used, and the techniques themselves are stochastic in nature. In addition, the optimization solver still needs to prove optimality of any solution (up to a tolerance level), which is itself a difficult task. For these reasons, there are no guarantees that providing an initial solution to an MIP solver will reduce computational time or effort. 

\subsubsection{Experiment Setup}
All simulations of the reconfiguration MILP were carried out with Gurobi, using the Yalmip interface. Different optimization solvers have different requirements for warm-start points, specifying whether the point must be feasible or not, and whether a full set of variables must be provided or not. For Gurobi MIPs the initial point must be feasible, but does not need to be complete for all variables. Grid-Si\textbf{PhyR} guarantees feasibility of any prediction for the equality constraints and a subset of the inequality constraints. Translating the Grid-Si\textbf{PhyR} prediction to a warm-start point can be done quite easily by simply omitting any variables which violate the inequality constraints. 

Three sets of simulations are carried. \textbf{Case 1:} the reconfiguration problem is solved without any warm-start. \textbf{Case 2:} the reconfiguration problem is solved by providing the optimal solution (from Gurobi) as the warm-start point to Gurobi. This provides a baseline comparison for the performance of warm-start on the reconfiguration problem. \textbf{Case 3:} the reconfiguration problem is solved by providing the Grid-Si\textbf{PhyR} prediction as the warm-start point to Gurobi. The optimization routine metrics are measured for each of these simulations. If the warm-start approach is effective for the reconfiguration MILP, it's expected that case 2 will significantly outperform both cases 1 and 3, and preferably case 3 will outperform case 1.

The simulations in case 2 were used to determine the set of variables to provide as a warm-start, denoted with the subscript `0'. When all variables $X_0 = \left[P^G\;Q^G\;V\;P_{ij}\;Q_{ij}\;y_{ij}\;z_{ij}\;z_{ji}\right]$ are provided as a warm-start point, the optimization solver is unable to identify it as a feasible solution. By testing different variables, the following were selected to provide as a warm-start: $X_0 = \left[P^G\;Q^G\;V\;y_{ij}\;z_{ij}\;z_{ji}\right]$. Any variables that violated inequality constraints were excluded.

\subsubsection{Performance metrics: optimization routine metrics}
The optimization routine metrics evaluate the performance of the Gurobi optimization solver when solving the MILP. It must be noted that the proposed PhyR-based framework is an end-to-end learning to solve the reconfiguration problem, and does not require an external solver. However, Gurobi is used to generate the data for the supervised framework. Additionally, a set of experiments are conducted to evaluate the PhyR-based framework as a warm-start technique for traditional optimization solvers. This experiment is evaluated based on the below optimization routine metrics.

\textbf{Solve time:} time in seconds to solve the optimization problem, through the YALMIP interface

\textbf{Node count:} number of branch-and-cut nodes explored in the most recent optimization

\textbf{Iteration count:} number of simplex iterations performed during the most recent optimization. Note that the reconfiguration problem is a linear program (LP) when the binary variables are fixed. Thus for every node in the branch-and-cut algorithm, the resulting optimization problem is an LP, efficiently solved using the simplex method.

\subsubsection{Experimental results for warm-start}

\textbf{Key takeaway:} Warm-start with the neural prediction can improve worst-case performance, but additional tuning is needed to further reduce solve time. Even then, there are no guarantees on optimizer performance. Warm-start for MIPs remains challenging.

Tables~\ref{tab:warmstart_BW} and \ref{tab:warmstart_TPC} present the results on the BW-33 and TPC-94 grids respectively. The key result from these results is that warm-start for the reconfiguration MIP is an ineffective technique, and may even result in poorer optimizer performance (along all three metrics). Interestingly, on average case 2 slightly increases solve time for the BW-33 grid while decreasing the node and iteration count, but is otherwise comparable. This difference across the metrics may be due to the size of the optimization problem and the relative time taken to reconstruct the full variable space and certify feasibility. These steps are not explicitly measured by the solver, but it has been noted by Gurobi developers that these actions can take considerable time. For the TPC-94 grid, case 2 improved optimizer performance on all metrics, by a larger margin than the BW-33 grid. Again, this may be due to the relative time taken in showing feasibility and optimality, versus searching the feasible space. Further, the search space for the TPC-94 grid is significantly larger than the BW-33 grid -- in topology alone, there are $2^{14}$ possible unique topologies for TPC-94 as compared to $2^7$ for BW-33. So although the optimality errors in neural prediction are comparable for the two grids, the warm-start in case 2 for the TPC-94 grid is more meaningful by starting closer to the optimal solution.

\begin{figure}[h]
    \centering
    \includegraphics[trim={8cm 0cm 8cm 0cm},clip,width=\linewidth]{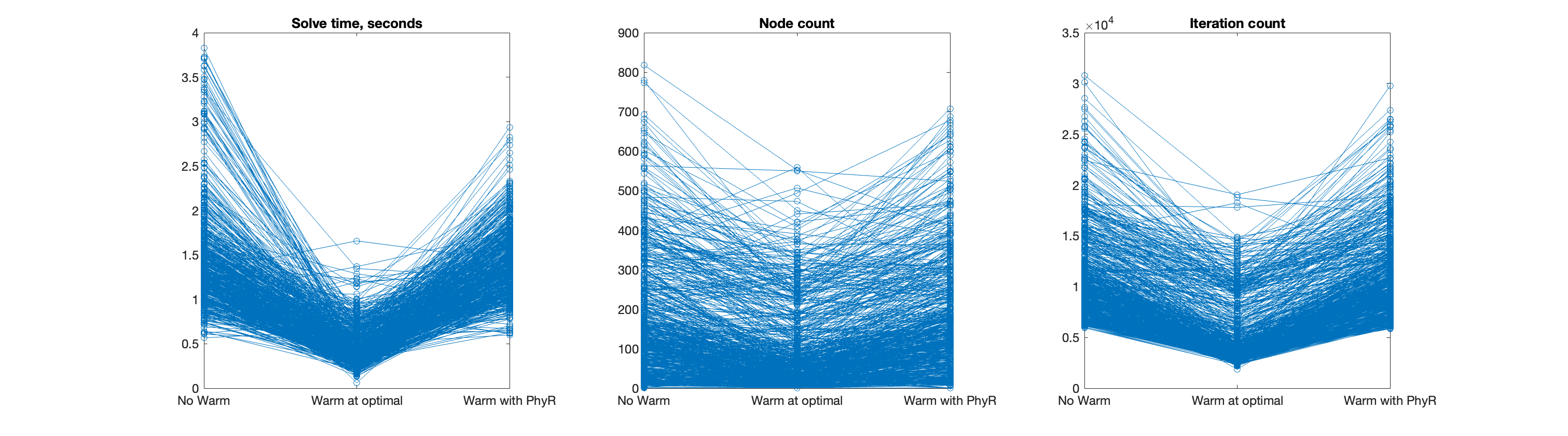}
    \caption{Warm-start results on the TPC-94 grid, plotting the optimization routine metrics. The warm-start experiment included 8640 data points, of which results for 700 randomly selected data points are plotted here.}
    \label{fig:results_warmstart}
\end{figure}

The performance of warm-start with case 3 is quite inconclusive. For the BW-33 grid, case 3 outperforms both case 1 and 2 on both mean and median solve times, which is unexpected. For the TPC-94 grid, case 3 underperforms against case 2 (expected) and is comparable to case 1. Notably the worst-case solve time in case 3 is significantly reduced over both cases 1 (desired) and 2 (unexpected). Figure~\ref{fig:results_warmstart} plots the warm-start metrics for all three cases, evaluated on the TPC-94 grid. The desired shape in the plot is an asymmetrical `V', where the left branch is longer than the right. This plot gives more insight into the performance of the different warm-start methods. Looking at the solve time, the desired shape is emerging, with case 2 outperforming the others and forming the vertex of the `V'. The shorter right side shows that while case 3 does not offer a reduction in solve time on average, it does permit a smaller in spread in solve times. This can be meaningful in applications where decisions must be made within a sensitive time window that is violated when using case 1, such as dynamic reconfiguration or electricity market clearing. Similar results are also seen for the other two metrics.  It is possible that improving the neural prediction performance will further improve the warm-start performance of case 3, with a lower bound in solve time provided by case 2.

\begin{table}[]
% \scriptsize
\centering
\caption{Warm-start results for the BW-33 grid}
\label{tab:warmstart_BW}
\begin{tabular}{llll}
\toprule
Metric & \begin{tabular}[c]{@{}c@{}}Case 1\\ Without warm-start\end{tabular} & \begin{tabular}[c]{@{}c@{}}Case 2\\ Warm-start at optimal\end{tabular} & \begin{tabular}[c]{@{}c@{}}Case 3\\ Warm-start at Grid-Si\textbf{PhyR} prediction\end{tabular} \\
\cmidrule{2-4}
Solve time & \begin{tabular}[c]{@{}l@{}}Mean: 0.2015\\ Median: 0.1586\\ Max: 0.7534\end{tabular} & \begin{tabular}[c]{@{}l@{}}Mean: 0.2057\\ Median: 0.1661\\ Max: 0.8775\end{tabular} & \begin{tabular}[c]{@{}l@{}}Mean: 0.1753\\ Median: 0.1371\\ Max: 1.0414\end{tabular} \\
Node count & \begin{tabular}[c]{@{}l@{}}Mean: 2.2094\\ Median: 1\end{tabular} & \begin{tabular}[c]{@{}l@{}}Mean: 1.8233\\ Median: 1\end{tabular} & \begin{tabular}[c]{@{}l@{}}Mean: 2.2215\\ Median: 1\end{tabular} \\
Iteration count & \begin{tabular}[c]{@{}l@{}}Mean: 2781\\ Median: 2629\end{tabular} & \begin{tabular}[c]{@{}l@{}}Mean: 2770\\ Median: 2656\end{tabular} & \begin{tabular}[c]{@{}l@{}}Mean: 2796\\ Median: 2663\end{tabular} \\
\bottomrule
\end{tabular}
\end{table}

\begin{table}[]
% \scriptsize
\centering
\caption{Warm-start results for the TPC-94 grid}
\label{tab:warmstart_TPC}
\begin{tabular}{llll}
\toprule
Metric & \begin{tabular}[c]{@{}c@{}}Case 1\\ Without warm-start\end{tabular} & \begin{tabular}[c]{@{}c@{}}Case 2\\ Warm-start at optimal\end{tabular} & \begin{tabular}[c]{@{}c@{}}Case 3\\ Warm-start at Grid-Si\textbf{PhyR} prediction\end{tabular} \\
\cmidrule{2-4}
Solve time & \begin{tabular}[c]{@{}l@{}}Mean: 1.4694\\ Median: 1.3322\\ Max: 6.6272\end{tabular} & \begin{tabular}[c]{@{}l@{}}Mean: 0.4516\\ Median: 0.3993\\ Max: 8.0498\end{tabular} & \begin{tabular}[c]{@{}l@{}}Mean: 1.4633\\ Median: 1.4157\\ Max: 3.4846\end{tabular} \\
Node count & \begin{tabular}[c]{@{}l@{}}Mean: 182\\ Median: 137\end{tabular} & \begin{tabular}[c]{@{}l@{}}Mean: 94.95\\ Median: 51\end{tabular} & \begin{tabular}[c]{@{}l@{}}Mean: 182.4\\ Median: 140\end{tabular} \\
Iteration count & \begin{tabular}[c]{@{}l@{}}Mean: 11405\\ Median: 10260\end{tabular} & \begin{tabular}[c]{@{}l@{}}Mean: 5290\\ Median: 3605\end{tabular} & \begin{tabular}[c]{@{}l@{}}Mean: 11453\\ Median: 10352\end{tabular} \\
\bottomrule
\end{tabular}
\end{table}

% old solve time for Vmin fixed: Mean: 0.6017\\ Median: 0.4323\\ Max: 4.0675

\section{Details on hyperparameter tuning} \label{app:hyperparam_tuning}
The following parameters were kept fixed for all neural network models across all experiments, based on a initial experimentation to ensure convergence and stability.
\begin{itemize}
    \item Epochs: 1500 for BW-33, and 2500 for TPC-94
    \item Batch size: 200
    \item Depth of network: 2 hidden layers; hidden layer size is kept uniform for both hidden layers
    \item Learning rate: $\left\{0.001, 0.0005, 0.0001 \right\}$
\end{itemize}
Based on the universal approximation theorem it is known that a neural network with a single hidden layer can represent any linear function, and one with two hidden layers can represent any arbitrary nonlinear function. While the underlying power flow constraints are linear (the reconfiguration problem is using the Linear DistFlow model), the presence of binary variables introduces a nonlinearity to the overall problem. Thus a network with 2 hidden layers is the smallest network that can be expected to perform well. Initial experiments were done to compare networks with one and two hidden layers, and corroborate this statement. A small number of initial experiments were also done to increase the number of hidden layers or adding 30\% dropout on each layer of the neural network. Overall these architectures provide minimal improvement in performance, and further tuning was not done.
% Notably, by having 2 hidden layers the performance improvement is more substantial for the the narrow networks with $N_{nn}=\{5,25\}$ neurons per layer. 

The key results from the hyperparameter tuning are summarized here:
\begin{itemize}
    \item Neural network width for BW-33 is 5 neurons per hidden layer, and for TPC-94 is 300 neurons per hidden layer. This captures the tradeoff between training speed and prediction performance on optimality and feasibility metrics
    \item InSi-based methods require lower learning rates (1e-4 versus 1e-3)
    \item Datasets with real load data (DS2 and DS5) require lower learning rates (1e-4 vs 1e-3)
\end{itemize}

Select results from the hyperparameter tuning are presented next.

\begin{figure}[h]
    \centering
    \includegraphics[width=0.65\linewidth]{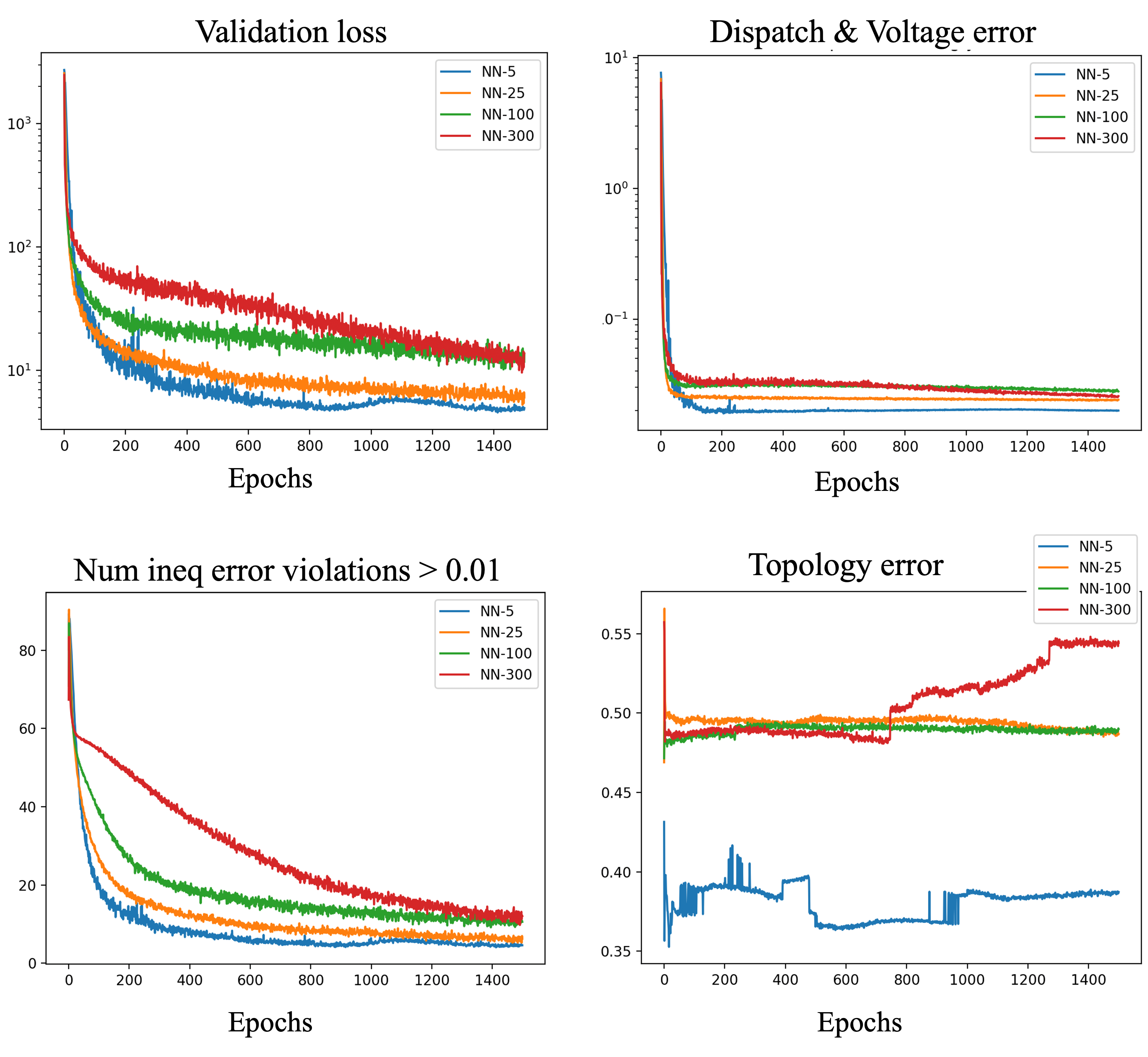}
    \caption{Tuning the hidden layer size for the BW-33 grid for the Si\textbf{PhyR} method. The learning rate is set to be $lr=0.001$.}
    \label{fig:ablation_BW_PhyR_2HL_NNsize}
\end{figure}

\begin{figure}
    \centering
    \includegraphics[width=0.65\linewidth]{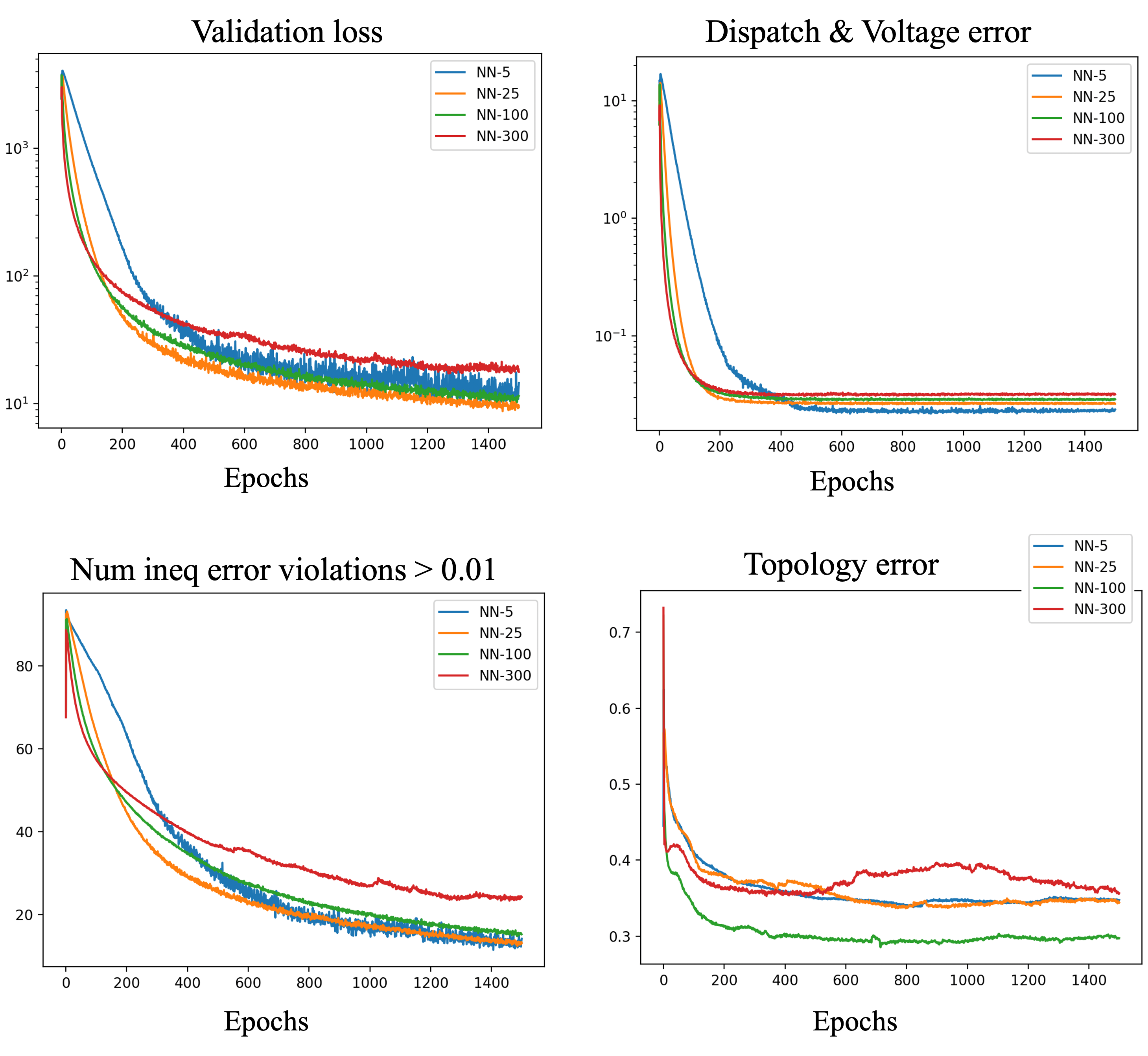}
    \caption{Tuning the hidden layer size for the BW-33 grid for the InSi method. The learning rate is set to be $lr=0.0001$.}
    \label{fig:ablation_BW_InSi_2HL_NNsize}
\end{figure}

Fig.~\ref{fig:ablation_BW_PhyR_2HL_NNsize} shows the validation results during neural training of Si\textbf{PhyR} with two hidden layers across different depths $N_{nn}$. Figure~\ref{fig:ablation_BW_InSi_2HL_NNsize} shows similar results for the InSi method. It must be noted that the InSi method requires a lower learning rate than Si\textbf{PhyR} (using $lr=0.0001$ for InSi vs. $lr=0.001$ for \textbf{PhyR}). Overall the training plots for both Si\textbf{PhyR} and InSi methods are comparable. The faster convergence in validation loss of Si\textbf{PhyR} may be attributed to the higher learning rate. The jumping behaviour in the topology error plot of Si\textbf{PhyR} as compared to the smoother plot of the InSi method can be explained by the nature of the physics-informed rounding: the Si\textbf{PhyR} method enforces integer values for switch status predictions, so a change in the topology prediction may result in step-like behaviour of the corresponding error metric. In comparison, the InSi method does not enforce integer solutions, so the corresponding metric is not expected to have a step-like behaviour. Overall the prediction performance with $N_{nn}=5$ is reasonable across all three plots, providing a tradeoff between fast convergence, low error, and having a light-weight small network. This width is selected for the BW-33 network for all further testing.

\begin{figure}[h]
    \centering
    \includegraphics[width=0.65\linewidth]{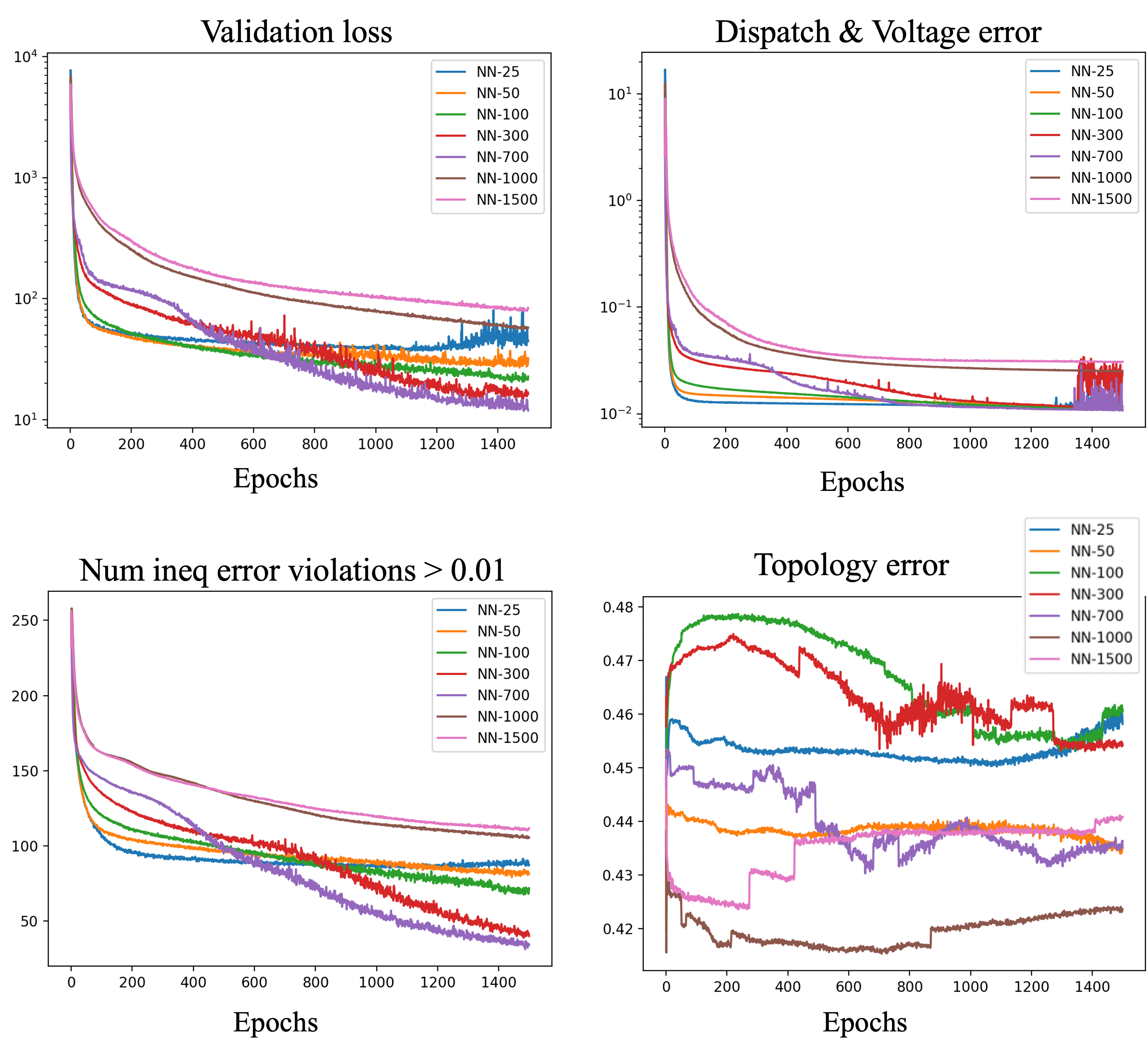}
    \caption{Tuning the hidden layer size for the TPC-94 grid for the Si\textbf{PhyR} method, on DS-1 with 9000 data points. The learning rate is set to be $lr=0.001$ for all $N_{nn}$ except $N_{nn}=\{1000,1500\}$ for which $lr=0.0001$.}
    \label{fig:ablation_TPC_PhyR_Nnn}
\end{figure}

Figure~\ref{fig:ablation_TPC_PhyR_Nnn} shows the hyperparameter tuning plots for Si\textbf{PhyR} on the TPC-94 grid. The prediction performance of $N_{nn}=\{300,700\}$ are best for the inequality violations, DispErr, and VoltErr, while the largest network $N_{nn}=1500$ performs the best on TopErr. This wider network however requires a lower learning rate, and so converges more slowly on the other performance metrics. The tradeoff between convergence time and overall prediction performance is shown here. The same experiment was conducted for the InSi method, showing similar performance across different $N_{nn}$. Of note, the InSi method required a lower learning rate ($lr = 0.0005$ or $lr=0.0001$). Overall the prediction performance with $N_{nn}=300$ is reasonable, and this width is selected for all further testing. It should be noted that the InSi method requires a lower learning rate than the PhyR method. The profile datasets (DS2 thru DS5) also require lower learning rates than the perturbed dataset DS1. 

\section{Grid reconfiguration problem} \label{app:network_MIP}
The grid reconfiguration problem using the Linear DistFlow model of the distribution grid \cite{Baran_reconfig_1989} is formulated as below.

% \begin{align}
%     \min_{\psi\in\mathcal{P}_x} & \; f_x(\psi) = \sum_{(i,j) \in \mathcal{T}_D} (P_{ij}^2 + P_{ji}^2 + Q_{ij}^2 + Q_{ji}^2)R_{ij} \label{eq:objectivefnc}\\
%     & P_j^G - P_j^L = \sum_{k: (j,k)}P_{jk} - \sum_{i:(i,j)}P_{ij}, \;  Q_j^G - Q_j^L = \sum_{k: (j,k)}Q_{jk} - \sum_{i:(i,j)}Q_{ij}, \quad \forall j\in \mathcal{B} \label{eq:pbalance} \\
%     & v_j - v_i = - 2(R_{ij}(P_{ij}-P_{ji}) + X_{ij}(Q_{ij}-Q_{ji})), \quad \forall (i,j)\in \mathcal{T}_\text{D}\setminus\mathcal{T}_\text{D}^{sw} \label{eq:Ohm0}\\
%     & z_{ij} + z_{ji} = 1, \quad \forall (i,j)\in \mathcal{T}_D \setminus \mathcal{T}_D^{sw} \label{eq:zconstLine} \\
%     & z_{ij} + z_{ji} = y_{ij}, \quad \forall (i,j)\in \mathcal{T}_D^{sw} \label{eq:zconstSwitch} \\
%     & \sum_{(i,j)\in \mathcal{T}_{D}^{sw}} y_{ij} = (N-1) - (M-M_{sw}) \label{eq:radial1}
% \end{align}

\begingroup
\footnotesize
\allowdisplaybreaks
\begin{subequations} \label{eq:app_reconfig}
\begin{align}
    \min_{\psi} \quad & f_x(\psi) \label{eq:app_objectivefnc}\\
    \textrm{s.t.} \quad 
    & v_j - v_i = - 2(R_{ij}(P_{ij}-P_{ji}) + X_{ij}(Q_{ij}-Q_{ji})) & \forall (i,j)\in \mathcal{T}_\text{D}\setminus\mathcal{T}_\text{D}^{sw} \label{eq:app_Ohm0}\\
    & v_j - v_i \leq - 2(R_{ij}(P_{ij}-P_{ji}) + X_{ij}(Q_{ij}-Q_{ji})) + M(1-y_{ij}) & \forall (i,j)\in \mathcal{T}_\text{D}^{sw} \label{eq:app_Ohm1}\\
    & v_j - v_i \geq - 2(R_{ij}(P_{ij}-P_{ji}) + X_{ij}(Q_{ij}-Q_{ji})) - M(1-y_{ij}) & \forall (i,j)\in \mathcal{T}_\text{D}^{sw} \label{eq:app_Ohm2}\\
    & P_j^G - P_j^L = \sum_{k: (j,k)}P_{jk} - \sum_{i:(i,j)}P_{ij}  & \forall j\in \mathcal{B} \label{eq:app_pbalance}\\
    & Q_j^G - Q_j^L = \sum_{k: (j,k)}Q_{jk} - \sum_{i:(i,j)}Q_{ij}  & \forall j\in \mathcal{B}\label{eq:app_qbalance} \\
    % & z_{i,j} \geq 0 & \forall (i,j),(j,i)\in \mathcal{T}_\text{D} \label{eq:app_zpositive} \\
    & z_{ij} + z_{ji} = 1  & \forall (i,j)\in \mathcal{T}_D \setminus \mathcal{T}_D^{sw} \label{eq:app_zconstLine} \\
    & z_{ij} + z_{ji} = y_{ij} & \forall (i,j)\in \mathcal{T}_D^{sw} \label{eq:app_zconstSwitch} \\
    & \sum_{(i,j) \in \mathcal{T}_D^{sw}} y_{ij} = (N-1) - (M-M_{sw}) & \label{eq:app_radial1} \\
    & \sum_{j:(i,j)} z_{ij} + z_{ji} \geq 1 & \forall j\in \mathcal{B} \label{eq:app_radial2} \\
    & y_{ij} \in \{0,1\} & \forall (i,j)\in \mathcal{T}_D^{sw} \label{eq:app_yBinary} \\
    & z_{ij}, z_{ji} \in \{0,1\} & \forall (i,j)\in \mathcal{T}_D \label{eq:app_zBinary} \\
    & 0 \leq P_{ij} \leq Mz_{ij} & \forall (i,j),(j,i)\in \mathcal{T}_D \label{eq:app_Pexist} \\
    & 0 \leq Q_{ij} \leq Mz_{ij} & \forall (i,j),(j,i)\in \mathcal{T}_D \label{eq:app_Qexist} \\
    % & P_j^L = P_j^{L,0} &\forall j\in \mathcal{B} \label{eq:app_PLlim}\\
    % & Q_j^L = Q_j^{L,0} &\forall j\in \mathcal{B} \label{eq:app_QLlim}\\
    & \underline{P_j^G} \leq P_j^G \leq \overline{P_j^G} &\forall j\in \mathcal{B} \label{eq:app_PGlim}\\
    & \underline{Q_j^G} \leq Q_j^G \leq \overline{Q_j^G} &\forall j\in \mathcal{B} \label{eq:app_QGlim}\\
    % & \underline{P_j^L} \leq P_j^L \leq \overline{P_j^L} &\forall j\in \mathcal{B} \label{eq:PLlim}\\
    % & \underline{Q_j^L} \leq Q_j^L \leq \overline{Q_j^L} &\forall j\in \mathcal{B} \label{eq:QLlim}\\
    % & \underline{F}_{ij} \leq F_{ij} \leq \overline{F}_{ij} &\forall (i,j)\in \mathcal{T}_\text{D} \label{eq:global7}\\
    & \underline{v_j} \leq v_j \leq \overline{v_j} &\forall j\in \mathcal{B}\setminus j^\# \label{eq:app_vlim}\\
    & v_{j^\#} = 1 &\label{eq:app_vslack}
\end{align}
\end{subequations}
\endgroup
In Eq.~\ref{eq:app_reconfig} a general distribution grid is described as a graph $\Gamma(\mathcal{B},\mathcal{T}_D)$, where $\mathcal{B}$ is the set of N nodes and $\mathcal{T}_D\coloneqq\{(i,j)\}$ is the set of M edges. The subset of lines with switches $\mathcal{T}_D^{sw} \subset \mathcal{T}_D$ contains $M_{sw}$ switches which can be turned on/off to change the grid topology. The distribution grid is connected to the point of common coupling (PCC) of the distribution grid to the bulk transmission grid, denoted by node $j^\#$. The power variables are the real and reactive power loads at every node $P^L_j, Q^L_j$, the real and reactive power generation at every node $P^G_j,Q^G_j$, the squared magnitude of nodal voltages $v$, and directed power flow through the distribution lines $P_{ij}, P_{ji}, Q_{ij}, Q_{ji}$. The line parameters of resistance and reactance are denoted as $R_{ij}$ and $X_{ij}$ respectively. The grid topology variables are the switch status $y_{ij} \forall (i,j) \in \mathcal{T}_{D}^{sw}\subset \mathcal{T}_{D}$ where $y_{ij} = 1$ is closed and $y_{ij} = 0$ is open, and the selection of direction of power flow through each line $z_{ij}, z_{ji}$. These variables are binary variables.

The objective function~\eqref{eq:app_objectivefnc} approximates the electrical line losses which when minimized ensures efficient grid operations. Constraints \eqref{eq:app_Ohm0}-\eqref{eq:app_Ohm2} describe Ohm's law across all lines in the network, with equality constraints describing lines without switches, and inequality constraints describing lines with switches. Notably, the big-M relaxation is used to describe the conditional constraints describing Ohm's law for lines $(i,j) \in \mathcal{T}_D^{sw}$ -- i.e. Ohm's law must be binding when the switches are closed ($y_{ij} =1$) but does not apply when the switches are open ($y_{ij} =0$). Constraints \eqref{eq:app_pbalance} and \eqref{eq:app_qbalance} describe lossless power balance at every node for both real and reactive power.

Constraints \eqref{eq:app_zconstLine}-\eqref{eq:app_Qexist} describe the topology selection through switch-status, grid radiality and connectivity constraints, enforce binary constraints on the $y_{ij}, z_{ij}, z_{ji}$ variables, and restrict power flow through a line based on switch-status. In particular, constraint~\eqref{eq:app_radial1} restricts the number of closed switches in the grid so it is radial with $N-1$ total branches, where $L$ switches must be closed and the remaining $M_{sw}-L$ must be open. This constraint is central to the application of the physics-informed rounding routine. Constraint~\eqref{eq:app_radial2} enforces connectivity by requiring power to flow into or out of a node along at least one line. It should be noted that typical reconfiguration problem statements also include an arborescence constraint \cite{Taylor_reconfigModel}, either explicitly or implicitly in the formulation of the radiality constraint. However, the increasing penetration of DERs voids this assumption, and multiple generating sources (roots of the tree) must be permitted. We have relaxed this arborescence constraint in~\eqref{eq:app_radial2}. Various other mathematical formulations of radiality and connectivity constraints include constraints on the determinant of the branch-to-node incidence matrix or spanning tree constraints and other graph theoretic approaches \cite{Lei_radiality_2020, Wang_radiality_2020, Ahmadi_radiality_2015, Lavorato_radiality_2012}. However, many of these suffer from high computational requirements and additional complexity, and do not leverage the fact that grid connectivity can be ensured by power flow constraints under normal operation. Our formulation accounts for this.

Constraints \eqref{eq:app_PGlim}-\eqref{eq:app_QGlim} describe generator operating limits, and \eqref{eq:app_vlim}-\eqref{eq:app_vslack} describes grid voltage limits where the voltage at the PCC is assumed to be fixed at 1pu, as is common practice in power systems. 

\paragraph{Extension for no export limits:} Additional set of constraints described below can be added to describe ``no export'' limits on the PCC, where net generation excess of net load in the distribution grid cannot be injected into the transmission grid. This is necessary in regions where distribution grids are not permitted to export power to the bulk grid, or where the amount of power that can be exported is limited.
\begin{subequations}
\begin{align}
    & z_{ij^\#} = 0 & \text{(no power flow from grid through PCC)} \label{eq:app_zfeeder} \\
    & \sum_{j:(j^\#,j)} P_{j^\#,j} = P_{j^\#}^G  & \text{(no real power export)} \label{eq:app_Pfeeder} \\
    & \sum_{j:(j^\#,j)} Q_{j^\#,j} = Q_{j^\#}^G & \text{(no reactive power export)} \label{eq:app_Qfeeder}
\end{align}    
\end{subequations}

\paragraph{Extension for power outage conditions:} Extensions of the presented model can be made for fault conditions where an element of the grid has failed (ex. a tree knocks down a power line, or a generator fails). The connectivity constraint must then be relaxed and islanding of sections of the grid is permitted, wherein the single distribution grid is broken down into smaller grids which are not electrically connected to one another. It must be noted however, that this is an undesirable phenomenon and only happens in extreme cases of outages and system failure. 

\paragraph{Extension for different constraints and applications:} Further extensions to the reconfiguration problem include distinctions between hard and soft constraints, considering the optimal switch change order to go from topology A to topology B, and considering grid outage conditions and subsequent generator restart and load recovery. Note that soft constraints can include lines which can exceed thermal limits for short periods of time during a reconfiguration activity.

\paragraph{Objective function:} The MILP detailed above is solved to minimize an objective function $f_x(\psi)$. For a modern distribution grid with high DER penetration, various objectives are sought after by grid operators. Some such objectives include minimizing electrical line losses (maximizing grid efficiency), minimizing costs for power generation, minimizing congestion, improving voltage profiles across the distribution feeder, reducing peak power demand, ensuring reliability of service (ex. higher capacity margins for feeders and supply transformers), and balancing load. Depending on the types of switches in the grid, operators may also minimize the cost incurred by actuating switches. In general, these objectives can be formulated using a convex function, thus retaining the uniqueness and global optimality of the optimal power flow solution. Note that the presence of binary variables means the search space is not convex nor continuous. 
% f_x(\psi) = \sum_{(i,j) \in \mathcal{T}_D} (P_{ij}^2 + P_{ji}^2 + Q_{ij}^2 + Q_{ji}^2)R_{ij}

\section{Datasets} \label{app:datasets}
We present the datasets used in the evaluation of the Grid-Si\textbf{PhyR} framework. Notably, there are few good training datasets available for power system applications, and no known datasets for the grid reconfiguration problem. The creation of these datasets is a contribution of this research. There are three relevant parameters to each dataset: (i) network topology, line parameters, and location of switches; (ii) location, magnitude, and time-varying profile of loads; and (iii) location, capacity, and time-varying profile of distributed generation. Note that (i)-(iii) are the input to the reconfiguration problem, and determine the optimal configuration of the network. In the sections that follow, each network dataset and (i)-(iii) will be presented. The network data (i) are taken from literature, as are the location and magnitude of loads for a single time period. The time-varying load profiles (ii) and the solar resource locations and generation profile (iii) are developed as part of the dataset generation. Table~\ref{tab:caseinfo} summarizes key parameters of the two distribution grids (BW-33 and TPC-94) presented in detail below.

\begin{table}[h]
\centering
\caption{Size of reconfiguration problem for two canonical distribution grids}
\begin{tabular}{lll}
\toprule 
 & BW-33 & TPC-94 \\
 \cmidrule{2-3}
Number of nodes, $|\mathcal{B}|$ & 33 & 94 \\
Number of lines and switches, $|\mathcal{T}_\text{D}|, |\mathcal{T}_\text{D}^{sw}|$ & 30, 8 & 83, 14 \\
\begin{tabular}[c]{@{}l@{}}Number of Discrete variables - topological, $y_{ij}$\end{tabular} & 8 & 14 \\
\begin{tabular}[c]{@{}l@{}}Number of Discrete variables - power flow, $z_{ij}, z_{ji}$\end{tabular} & 74 & 194 \\
Number of Continuous vars & 248 & 671 \\
Number of Equality constraints & 134 & 369 \\
Number of Inequality constraints & 545 & 1465 \\
Size of Training data, $|x|$ & 64 & 186 \\
Size of Independent variables, $|z|$ & 195 & 510 \\
Size of Dependent variables, $|\varphi|$ & 134 & 369 \\
\bottomrule 
\end{tabular}
\label{tab:caseinfo}
\end{table}

\subsection{33-Node Baran-Wu Grid, BW-33}
\begin{figure}[!h] 
	\centering
	% 3 colume: 0.31 width, 5.4cm
	% 2 colume: 0.48 width, 8.4 / 7.8 cm.
	\subfloat[BW-33 distribution grid \label{fig:33node_feeder}]{\includegraphics[width=0.36\textwidth]{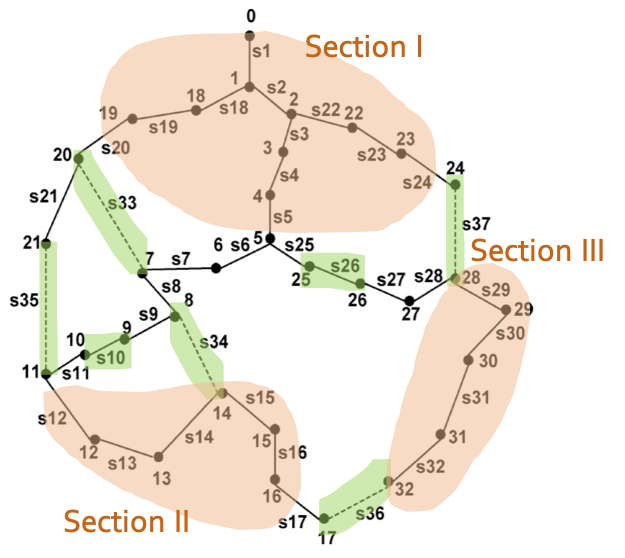}}
% 	\hspace{0.2cm}
	\subfloat[DD-U \label{fig:33node_DD-U}]{\includegraphics[width=0.33\textwidth]{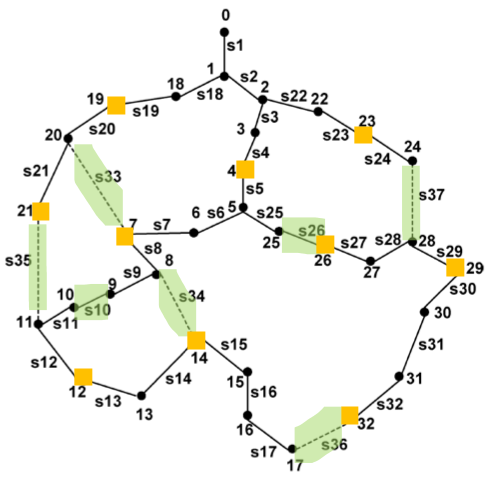}}
% 	\hspace{0.2cm}
	\subfloat[DD-I \label{fig:33node_DD-I}]{\includegraphics[width=0.33\textwidth]{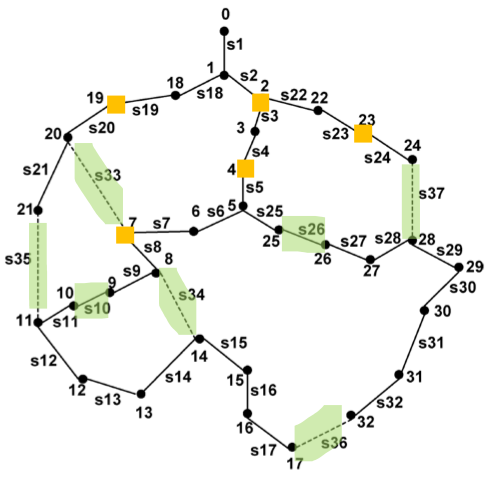}}
	\\
	\subfloat[DD-II \label{fig:33node_DD-II}]{\includegraphics[width=0.33\textwidth]{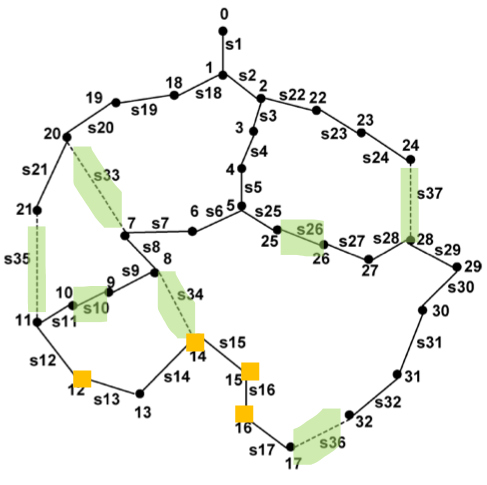}}
% 	\hspace{0.2cm}
	\subfloat[DD-III \label{fig:33node_DD-III}]{\includegraphics[width=0.33\textwidth]{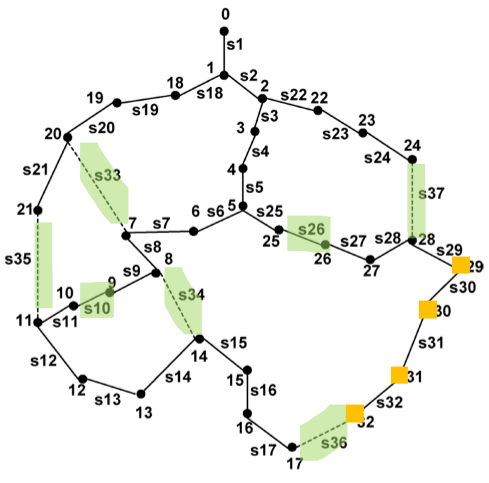}}
% 	\hspace{0.2cm}
	\subfloat[DD-II+III \label{fig:33node_DD-II_III}]{\includegraphics[width=0.33\textwidth]{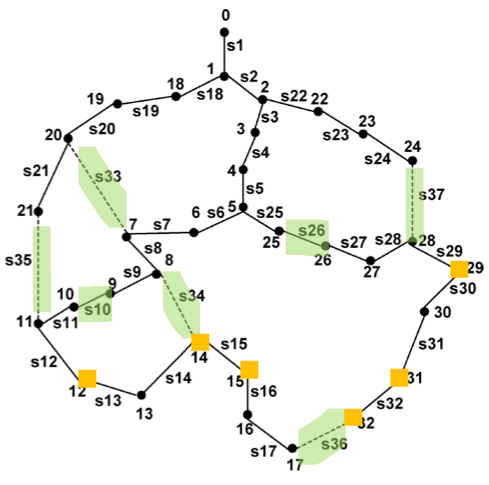}}
	\caption{BW-33 distribution grid from \cite{Baran_reconfig_1989}. The switches are highlighted in green, sections of the grid highlighted and labelled in orange, and location of community solar DERs noted in yellow squares.}
	\label{fig:33node_DD}
\end{figure}

The BW-33 grid presented in \cite{Baran_reconfig_1989} is a canonical grid used in the reconfiguration literature. The grid is very lossy, with losses up to 8\% of total load, and voltage profile violating voltage limits. These characteristics make the BW-33 grid an excellent test case for dynamic reconfiguration, with the objective function to reduce line losses. The network is shown in Fig.~\ref{fig:33node_feeder}. Grid data available in \cite{Baran_reconfig_1989} includes the grid topology and line parameters, as presented in Table~\ref{tab:line_params_33}, and location of loads and their nominal power demand (P and Q) for a single period as presented in Table~\ref{tab:load_data_33}. 

% \textcolor{red}{UPDATE FIGURE TO ADD SWITCH 4}

\subsubsection{Network Topology and Parameters}
The grid consists of 33 nodes ($N=33$), 37 lines ($M=37$) of which 5 are tie lines (normally open switches, NOS) and the remaining 32 are typically assumed to be sectionalizing switches (normally closed switches, NCS). Topology and line parameter data are presented in Table~\ref{tab:line_params_33}. To restrict the problem to a simpler test case, only a subset of the lines are considered switchable -- these include the 5 tie lines (numbered 33 to 37) and 3 NCS lines (line numbers 4, 10, and 26). These are highlighted in green in Fig.~\ref{fig:33node_feeder}, with dashed lines representing normally open switches and solid lines representing normally closed switches. 

\subsubsection{Load Location and Profiles}
The location of loads and their nominal power demand (P and Q) for a single period is presented in Table~\ref{tab:load_data_33}. To develop a diverse set of training data, the maximum load perturbation at each node is restricted to 70\% deviation from nominal value (i.e. $P'=\delta P, \delta=[0.3, 1.7]$). The power factor of the loads, which describes the relationship between the real and reactive power as $pf = P/(\sqrt{P^2 + Q^2})$, is kept constant to the pf in the nominal data. This is a common approach used in literature \cite{Fioretto_lagrangeduals,Zamzam_SmartGridComm2020,vpoor_opf_topologyreconfig}.

\subsubsection{Distributed Generation: Community Solar Dataset}
\begin{figure}[!h]
    \centering
    \includegraphics[width=0.5\textwidth]{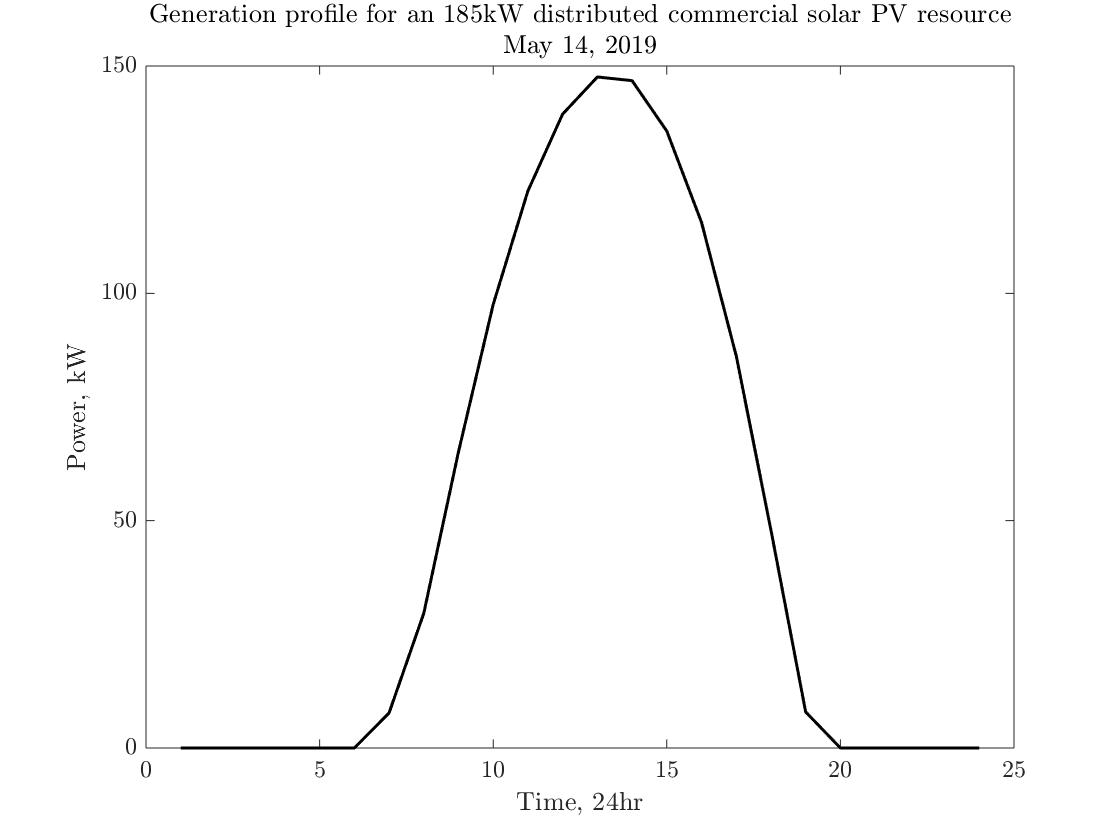}
    \caption{Sample solar PV generation profile for community PV facility, queried for May 14, 2019.}
    \label{fig:genProfile_solardistcomm}
\end{figure}

We add a range of community solar facilities (each <5MW), up to a penetration of 25.3\% of nameplate capacity to baseline load. This is a modest DER penetration compared to that which we would expect in the future grid, and reflects solar uptake now and over the next few years. We divide the grid into sections, based on the location of switches, and vary the location of community solar farms amongst these sections. We denote the distribution of these DERs (DD) as follows: (i) DD-U: uniform distribution of solar throughout the grid; (ii-iv) DD-I, DD-II, DD-III: all facilities are in Sections I, II, or III of the grid respectively; (v) DD-II+III: all facilities are in Sections II and III of the grid. The DD and location of each community solar facility is shown in Fig.~\ref{fig:33node_DD}, as indicated by the yellow squares. Different DDs are used to consider effect on grid reconfiguration, line losses, and voltage profiles. The location and nameplate capacity of each solar facility is provided in Table~\ref{tab:solar_locations_33}. Figure~\ref{fig:genProfile_solardistcomm} shows a sample 24-hour generation profile of a solar facility in the BW-33 dataset, at hourly intervals.

The solar generation data is taken from NREL's System Advisory Model (SAM) tool \cite{nrelSAM}. The data is of a 185kW distributed commercial solar PV facility, located in Phoenix, AZ, using the SunPower SPR-E19-310-COM module, and SMA America (STP 60-US-10, 400V) inverter. The DC to AC ratio is set to the default of 1.2. The desired array size is set to 220kWdc, giving a total AC capacity of 179.580kWac. All other parameters are left unchanged in the SAM setup.

\begin{longtable}{l|llll}
\caption{BW-33 grid topology data and line parameters} \label{tab:line_params_33}\\
% \begin{tabular}{lllll}
\toprule
Branch No. & Upstream Node & Downstream Node & R [ohm] & X [ohm] \\ \hline
1 & 1 & 2 & 0.0922 & 0.0470 \\
2 & 2 & 3 & 0.4930 & 0.2511 \\
3 & 3 & 4 & 0.3660 & 0.1864 \\
4 & 4 & 5 & 0.3811 & 0.1941 \\
5 & 5 & 6 & 0.8190 & 0.707 \\ 
6 & 6 & 7 & 0.1872 & 0.6188 \\ 
7 & 7 & 8 & 0.7114 & 0.2351 \\
8 & 8 & 9 & 1.030 & 0.7400 \\
9 & 9 & 10 & 1.0440 & 0.7400 \\
10 & 10 & 11 & 0.1966 & 0.0650 \\
11 & 11 & 12 & 0.3744 & 0.1238 \\
12 & 12 & 13 & 1.4680 & 1.1550 \\
13 & 13 & 14 & 0.5416 & 0.7129 \\
14 & 14 & 15 & 0.5910 & 0.5260 \\
15 & 15 & 16 & 0.7463 & 0.5450 \\
16 & 16 & 17 & 1.2890 & 1.7210 \\
17 & 17 & 18 & 0.7320 & 0.5740 \\
18 & 2 & 19 & 0.1640 & 0.1565 \\
19 & 19 & 20 & 1.5042 & 1.3554 \\
20 & 20 & 21 & 0.4095 & 0.4784 \\
21 & 21 & 22 & 0.7089 & 0.9373 \\
22 & 3 & 23 & 0.4512 & 0.3083 \\
23 & 23 & 24 & 0.8980 & 0.7091 \\
24 & 24 & 25 & 0.8960 & 0.7011 \\
25 & 6 & 26 & 0.2030 & 0.1034 \\
26 & 26 & 27 & 0.2842 & 0.1447 \\
27 & 27 & 28 & 1.0590 & 0.9337 \\
28 & 28 & 29 & 0.8042 & 0.7006 \\
29 & 29 & 30 & 0.5075 & 0.2585 \\
30 & 30 & 31 & 0.9744 & 0.9630 \\
31 & 31 & 32 & 0.3105 & 0.3619 \\
32 & 32 & 33 & 0.3410 & 0.5302 \\
33 & 8 & 21 & 2.00 & 2.00 \\
34 & 9 & 15 & 2.00 & 2.00 \\
35 & 12 & 22 & 2.00 & 2.00 \\
36 & 18 & 33 & 0.500 & 0.500 \\
37 & 25 & 29 & 0.500 & 0.500 \\
\bottomrule
% \end{tabular}
\end{longtable}

\begin{longtable}{lll|lll|lll}
% \begin{table}[!h]
\caption{BW-33 grid load data} \label{tab:load_data_33} \\
% \begin{tabular}{lll|lll|lll}
\toprule 
$j$ & $P^L$ [kW] & $Q^L$ [kVAR] & $j$ & $P^L$ [kW] & $Q^L$ [kVAR] & $j$ & $P^L$ [kW] & $Q^L$ [kVAR]\\ \hline
2 & 100 & 60 & 13 & 60 & 35 & 24 & 420 & 200 \\
3 & 90 & 40 & 14 & 120 & 80 & 25 & 420 & 200 \\
4 & 120 & 80 & 15 & 60 & 10 & 26 & 60 & 25 \\
5 & 60 & 30 & 16 & 60 & 20 & 27 & 60 & 25 \\
6 & 60 & 20 & 17 & 60 & 20 & 28 & 60 & 20 \\
7 & 200 & 100 & 18 & 90 & 40 & 29 & 120 & 70 \\
8 & 200 & 100 & 19 & 90 & 40 & 30 & 200 & 600 \\
9 & 60 & 20 & 20 & 90 & 40 & 31 & 150 & 70 \\
10 & 60 & 20 & 21 & 90 & 40 & 32 & 210 & 100 \\
11 & 45 & 30 & 22 & 90 & 40 & 33 & 60 & 40 \\
12 & 60 & 35 & 23 & 90 & 50 &  &  & \\
\bottomrule
\end{longtable}
% \end{tabular}
% \end{table}

\begin{longtable}{ll|ll|ll|ll|ll}
\caption{Locations and capacity of community solar facilities under each DD. The generating capacity is in kW} \label{tab:solar_locations_33} \\
\toprule 
\multicolumn{2}{c|}{DD-U} & \multicolumn{2}{c|}{DD-I} & \multicolumn{2}{c|}{DD-II} & \multicolumn{2}{c|}{DD-III} & \multicolumn{2}{l}{DD-II+III} \\ \hline
$j$ & $\overline{P}^G$ & $j$ & $\overline{P}^G$ & $j$ & $\overline{P}^G$ & $j$ & $\overline{P}^G$ & $j$ & $\overline{P}^G$ \\ \hline
4 & 60 & 2 & 185 & 12 & 300 & 29 & 300 & 12 & 170 \\
7 & 100 & 4 & 160 & 14 & 160 & 30 & 160 & 14 & 100 \\
12 & 120 & 7 & 200 & 15 & 200 & 31 & 200 & 15 & 180 \\
14 & 80 & 19 & 185 & 16 & 280 & 32 & 280 & 29 & 160 \\
19 & 110 & 23 & 210 &  &  &  &  & 31 & 190 \\
23 & 80 &  &  &  &  &  &  & 32 & 140 \\
21 & 110 &  &  &  &  &  &  &  &  \\
26 & 70 &  &  &  &  &  &  &  &  \\
29 & 60 &  &  &  &  &  &  &  &  \\
32 & 150 &  &  &  &  &  &  &  & \\
\bottomrule
\end{longtable}

\subsubsection{Dataset}
The BW-33 grid dataset consists of 8760 data points, using hourly load and solar generation over a year.

\subsection{94-Node Distribution Grid, TPC-94}
\begin{figure}[tbhp]
    \centering
    \includegraphics[width=0.7\linewidth]{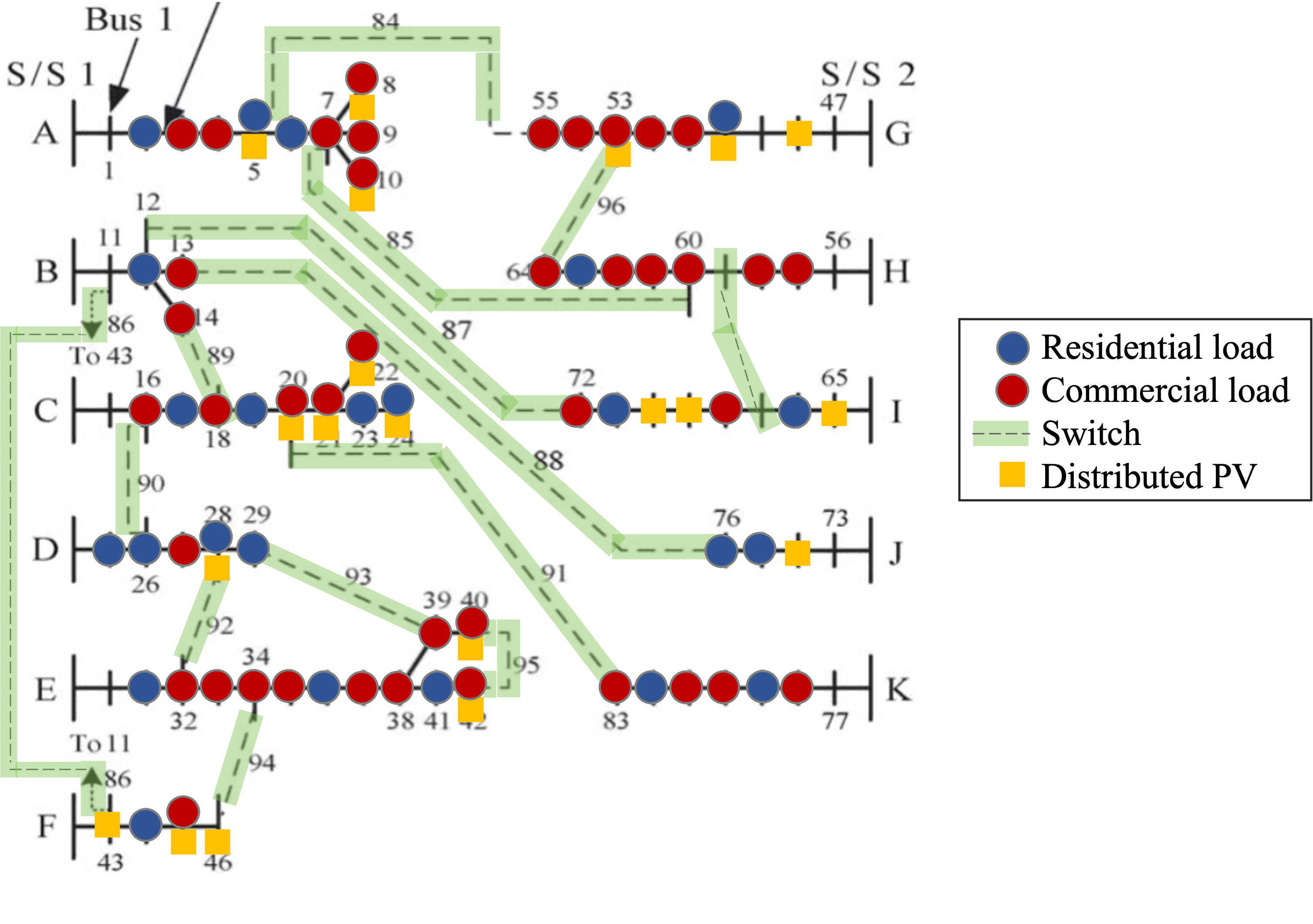} % width=0.8\textwidth
    \caption{94-node distribution grid from \cite{Su_2005_83dataset}. The switches are highlighted in green. The location of residential and commercial loads are noted by blue and red circles respectively. The location of community solar DERs are noted in yellow squares (for the S1 dataset). In the L1 and L2 dataset, all commercial loads (red) are assumed to follow residential profiles (blue). In the L3 dataset, the residential and commercial loads are located as per this figure. Note that the selection of locations of residential loads, commercial loads, and DERs, and their corresponding load/generation profiles are not part of the original network data. }
    \label{fig:83node_diagram}
\end{figure}

The TPC-94 grid is presented in \cite{Su_2005_83dataset}, and is a practical distribution network of Taiwan Power Company. The data available in \cite{Su_2005_83dataset}. The network consists of 11 feeders which are able to share load and generation by using tie line switches. The grid was modified by adding different loads and generation. The resulting grid is shown in Fig.~\ref{fig:83node_diagram}, with the locations of switches, residential and commercial loads, and distributed PV installations marked.

\subsubsection{Network Topology and Parameters}
The grid consists of 94 nodes ($N=94$), of which 11 are the T-D substations for feeder heads A thru K, and 83 are the remaining nodes in the network. The grid has 97 lines ($M=97$) of which 14 are tie lines (NOS) and the remaining 83 are typically assumed to be sectionalizing switches (NCS). Topology and line parameter data are presented in Table~\ref{tab:line_params_83}. To restrict the problem to that of dynamic reconfiguration during nominal operation (i.e. full network is connected), only the tie lines are considered as switches: these include the original 13 tie lines and a new tie line introduced between nodes 59 and 67. This tie line is introduced to allow the 11 feeders to be connected to one another; in the original network, there are two sub-networks consisting of feeders A, G and H, and the remaining feeders B, C, D, E, F, I, J, and K. The new tie line connects feeders H and I.

\begin{longtable}{l|llll}
\caption{TPC-94 grid topology data and line parameters} \label{tab:line_params_83}\\
% \begin{tabular}{l|llll}
\toprule
Branch No. & Upstream Node & Downstream Node & R [ohm] & X [ohm] \\ \hline
    1  &  1  &  12  &  0.0085  &  0.0290 \\ 
     2  &  12  &  13  &  0.0092  &  0.0188 \\ 
     3  &  13  &  14  &  0.0103  &  0.0212 \\ 
     4  &  14  &  15  &  0.0040  &  0.0082 \\ 
     5  &  15  &  16  &  0.0092  &  0.0188 \\ 
     6  &  16  &  17  &  0.0017  &  0.0035 \\ 
     7  &  17  &  18  &  0.0018  &  0.0060 \\ 
     8  &  18  &  19  &  0.0046  &  0.0094 \\ 
     9  &  18  &  20  &  0.0103  &  0.0212 \\ 
     10  &  18  &  21  &  0.0046  &  0.0094 \\ 
     11  &  2  &  22  &  0.0034  &  0.0071 \\ 
     12  &  22  &  23  &  0.0149  &  0.0303 \\ 
     13  &  23  &  24  &  0.0011  &  0.0024 \\ 
     14  &  23  &  25  &  0.0034  &  0.0071 \\ 
     15  &  3  &  26  &  0.0050  &  0.0169 \\ 
     16  &  26  &  27  &  0.0023  &  0.0047 \\ 
     17  &  27  &  28  &  0.0023  &  0.0047 \\ 
     18  &  28  &  29  &  0.0069  &  0.0141 \\ 
     19  &  29  &  30  &  0.0017  &  0.0035 \\ 
     20  &  30  &  31  &  0.0074  &  0.0153 \\ 
     21  &  31  &  32  &  0.0103  &  0.0212 \\ 
     22  &  32  &  33  &  0.0069  &  0.0141 \\ 
     23  &  32  &  34  &  0.0086  &  0.0176 \\ 
     24  &  34  &  35  &  0.0057  &  0.0118 \\ 
     25  &  4  &  36  &  0.0025  &  0.0084 \\ 
     26  &  36  &  37  &  0.0046  &  0.0094 \\ 
     27  &  37  &  38  &  0.0109  &  0.0223 \\ 
     28  &  38  &  39  &  0.0021  &  0.0072 \\ 
     29  &  39  &  40  &  0.0057  &  0.0118 \\ 
     30  &  5  &  41  &  0.0086  &  0.0173 \\ 
     31  &  41  &  42  &  0.0057  &  0.0118 \\ 
     32  &  42  &  43  &  0.0057  &  0.0118 \\ 
     33  &  43  &  44  &  0.0011  &  0.0024 \\ 
     34  &  44  &  45  &  0.0074  &  0.0153 \\ 
     35  &  45  &  46  &  0.0023  &  0.0047 \\ 
     36  &  46  &  47  &  0.0218  &  0.0447 \\ 
     37  &  47  &  48  &  0.0017  &  0.0035 \\ 
     38  &  48  &  49  &  0.0017  &  0.0035 \\ 
     39  &  49  &  50  &  0.0034  &  0.0071 \\ 
     40  &  50  &  51  &  0.0092  &  0.0188 \\ 
     41  &  49  &  52  &  0.0086  &  0.0176 \\ 
     42  &  52  &  53  &  0.0092  &  0.0188 \\ 
     43  &  6  &  54  &  0.0021  &  0.0072 \\ 
     44  &  54  &  55  &  0.0017  &  0.0035 \\ 
     45  &  55  &  56  &  0.0057  &  0.0118 \\ 
     46  &  56  &  57  &  0.0103  &  0.0212 \\ 
     47  &  7  &  58  &  0.0106  &  0.0362 \\ 
     48  &  58  &  59  &  0.0029  &  0.0059 \\ 
     49  &  59  &  60  &  0.0029  &  0.0059 \\ 
     50  &  60  &  61  &  0.0017  &  0.0035 \\ 
     51  &  61  &  62  &  0.0034  &  0.0071 \\ 
     52  &  62  &  63  &  0.0017  &  0.0035 \\ 
     53  &  63  &  64  &  0.0034  &  0.0071 \\ 
     54  &  64  &  65  &  0.0023  &  0.0047 \\ 
     55  &  65  &  66  &  0.0057  &  0.0118 \\ 
     56  &  8  &  67  &  0.0099  &  0.0338 \\ 
     57  &  67  &  68  &  0.0235  &  0.0482 \\ 
     58  &  68  &  69  &  0.0023  &  0.0047 \\ 
     59  &  69  &  70  &  0.0018  &  0.0060 \\ 
     60  &  70  &  71  &  0.0017  &  0.0035 \\ 
     61  &  71  &  72  &  0.0011  &  0.0024 \\ 
     62  &  72  &  73  &  0.0046  &  0.0094 \\ 
     63  &  73  &  74  &  0.0103  &  0.0212 \\ 
     64  &  74  &  75  &  0.0011  &  0.0036 \\ 
     65  &  9  &  76  &  0.0021  &  0.0072 \\ 
     66  &  76  &  77  &  0.0074  &  0.0153 \\ 
     67  &  77  &  78  &  0.0053  &  0.0181 \\ 
     68  &  78  &  79  &  0.0096  &  0.0326 \\ 
     69  &  79  &  80  &  0.0021  &  0.0072 \\ 
     70  &  80  &  81  &  0.0032  &  0.0109 \\ 
     71  &  81  &  82  &  0.0025  &  0.0084 \\ 
     72  &  82  &  83  &  0.0011  &  0.0023 \\ 
     73  &  10  &  84  &  0.0142  &  0.0483 \\ 
     74  &  84  &  85  &  0.0014  &  0.0048 \\ 
     75  &  85  &  86  &  0.0025  &  0.0084 \\ 
     76  &  86  &  87  &  0.0021  &  0.0072 \\ 
     77  &  11  &  88  &  0.0110  &  0.0374 \\ 
     78  &  88  &  89  &  0.0057  &  0.0193 \\ 
     79  &  89  &  90  &  0.0021  &  0.0072 \\ 
     80  &  90  &  91  &  0.0057  &  0.0115 \\ 
     81  &  91  &  92  &  0.0057  &  0.0115 \\ 
     82  &  92  &  93  &  0.0040  &  0.0082 \\ 
     83  &  93  &  94  &  0.0137  &  0.0282 \\ 
     84  &  16  &  66  &  0.0057  &  0.0118 \\ 
     85  &  18  &  71  &  0.0057  &  0.0118 \\ 
     86  &  22  &  54  &  0.0057  &  0.0118 \\ 
     87  &  23  &  83  &  0.0149  &  0.0306 \\ 
     88  &  24  &  87  &  0.0200  &  0.0411 \\ 
     89  &  25  &  29  &  0.0235  &  0.0473 \\ 
     90  &  27  &  37  &  0.0040  &  0.0082 \\ 
     91  &  31  &  94  &  0.0034  &  0.0071 \\ 
     92  &  39  &  43  &  0.0023  &  0.0047 \\ 
     93  &  40  &  50  &  0.0034  &  0.0071 \\ 
     94  &  45  &  57  &  0.0011  &  0.0024 \\ 
     95  &  51  &  53  &  0.0086  &  0.0176 \\ 
     96  &  64  &  75  &  0.0017  &  0.0035 \\ 
     97  &  70  &  78  &  0.0077  &  0.0158 \\
     \bottomrule
\end{longtable}

\begin{longtable}{lll|lll|lll}
% \begin{table}[!h]
\caption{TPC-94 grid load data} \label{tab:load_data_83} \\
% \begin{tabular}{lll|lll|lll}
\toprule
$j$ & $P^L$ [kW] & $Q^L$ [kVAR] & $j$ & $P^L$ [kW] & $Q^L$ [kVAR] & $j$ & $P^L$ [kW] & $Q^L$ [kVAR]\\ \hline
13  &  100  &  50  &  14  &  300  & 200  &  15  &  350  & 250 \\ 
     16  &  220  &  100  &  17  &  1100  & 800  &  18  &  400  & 320 \\ 
     19  &  300  &  200  &  20  &  300  & 230  &  21  &  300  & 260 \\ 
     23  &  1200  &  800  &  24  &  800  & 600  &  25  &  700  & 500 \\ 
     27  &  300  &  150  &  28  &  500  & 350  &  29  &  700  & 400 \\ 
     30  &  1200  &  1000  &  31  &  300  & 300  &  32  &  400  & 350 \\ 
     33  &  50  &  20  &  34  &  50  & 20  &  35  &  50  & 10 \\ 
     36  &  50  &  30  &  37  &  100  & 60  &  38  &  100  & 70 \\ 
     39  &  1800  &  1300  &  40  &  200  & 120  &  42  &  1800  & 1600 \\ 
     43  &  200  &  150  &  44  &  200  & 100  &  45  &  800  & 600 \\ 
     46  &  100  &  60  &  47  &  100  & 60  &  48  &  20  & 10 \\ 
     49  &  20  &  10  &  50  &  20  & 10  &  51  &  20  & 10 \\ 
     52  &  200  &  160  &  53  &  50  & 30  &  55  &  30  & 20 \\ 
     56  &  800  &  700  &  57  &  200  & 150  &  61  &  200  & 160 \\ 
     62  &  800  &  600  &  63  &  500  & 300  &  64  &  500  & 350 \\ 
     65  &  500  &  300  &  66  &  200  & 80  &  68  &  30  & 20 \\ 
     69  &  600  &  420  &  71  &  20  & 10  &  72  &  20  & 10 \\ 
     73  &  200  &  130  &  74  &  300  & 240  &  75  &  300  & 200 \\ 
     77  &  50  &  30  &  79  &  400  & 360  &  82  &  2000  & 1500 \\ 
     83  &  200  &  150  &  86  &  1200  & 950  &  87  &  300  & 180 \\ 
     89  &  400  &  360  &  90  &  2000  & 1300  &  91  &  200  & 140 \\ 
     92  &  500  &  360  &  93  &  100  & 30  &  94  &  400  & 360 \\
     \bottomrule
\end{longtable}
% \end{tabular}
% \end{table}

\subsubsection{Load Location and Profiles}
The location of loads and their nominal power demand (P and Q) for a single period is presented in Table~\ref{tab:load_data_83}. In the TPC-94 grid dataset, three different load datasets are generated. 

\begin{figure}
    \centering
    \includegraphics[width=\linewidth]{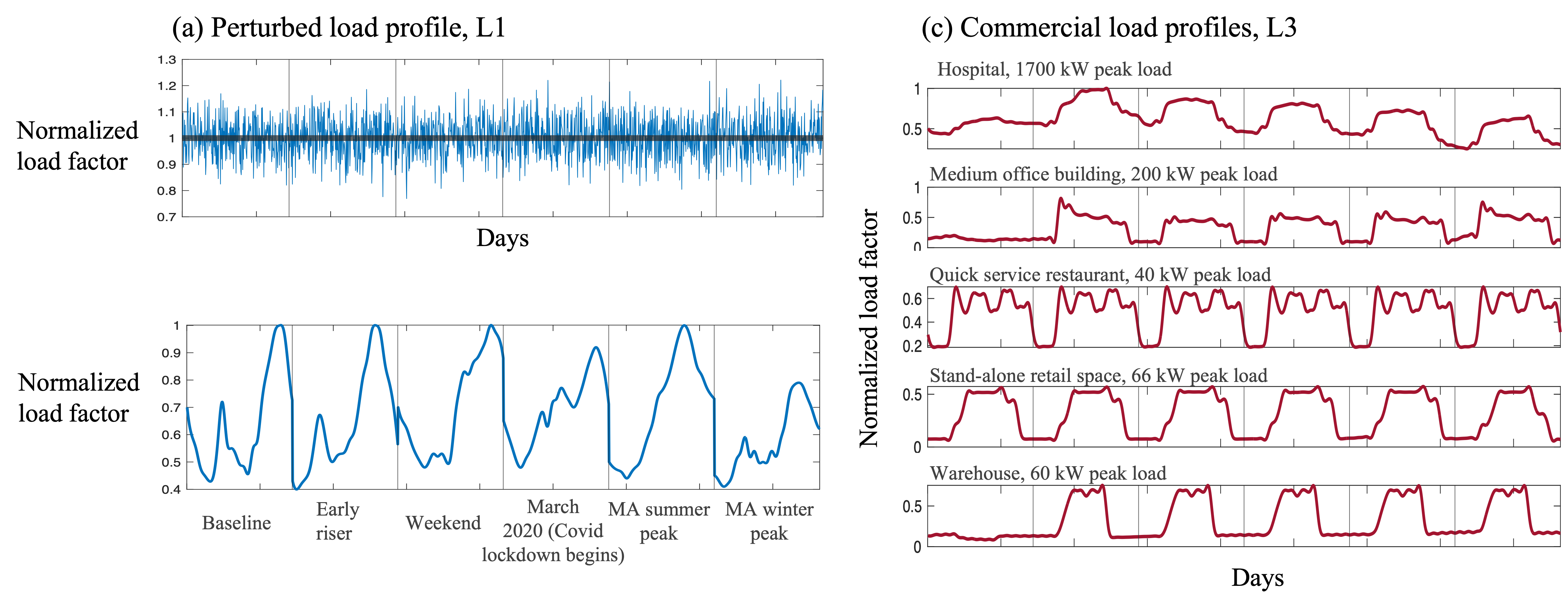}
    \caption{Load profiles for the perturbed dataset (L1), residential dataset (L2), and commercial load profiles for the mixed dataset (L3).}
    \label{fig:loadprofiles}
\end{figure}

\textbf{Perturbed dataset, L1:} The first load dataset generates random load perturbations, similar to the BW-33 grid. The maximum load perturbation at each node is restricted to 70\% deviation from nominal value (i.e. $P'=\delta P, \delta=[0.3, 1.7]$). The power factor of the loads, which describes the relationship between the real and reactive power as $pf = P/(\sqrt{P^2 + Q^2})$, is kept constant to the pf in the nominal data. Sample data over 6 days is shown in Fig.\ref{fig:loadprofiles}(a).

\textbf{Residential dataset, L2:} The second load dataset assumes all loads to be residential loads, which follow one of six profiles. The power factor of the loads is kept constant to the pf in the nominal data. The six load profiles are shown in Fig.\ref{fig:loadprofiles}(b). The six residential load profiles are described below:
\begin{itemize}
    \item Nominal: Typical residential load profile from \cite{Evans_loaddata_2003} exhibiting the characteristic bimodal distribution for residential load. The first mode occurs in the morning hours between 6-9am, when residential customers wake up and begin consuming electricity. The second mode occurs in the evening, after 6pm, when residents are returning home from work. 
    \item Early riser: A variation of the nominal profile with both modes shifted earlier in the day. This profile represents a residential customer which wakes earlier in the morning, and retires earlier in the evening.
    \item Weekend/Late night: A variation of the nominal profile for weekend residential consumption. This profile has higher electricity usage throughout the day, with a unimodal distribution of afternoon and evening consumption.
    \item Early-Covid19 (March and April 2020): The residential load profile shifted visibly during the Covid-19 pandemic, in particular during the early months of March and April. Analysis on residential electricity demand for the province on Ontario (Canada) show significant increase in daily electricity consumption, with a delayed morning peak and a higher maximum peak in the evening hours. The overnight consumption remains typical to the nominal residential profile. Residential load profile reported by the Ontario Independent Electricity System Operator (IESO) is presented in \cite{Abdeen_loaddata_covid}, and used to inform the load profile.
    \item Massachusetts peak summer and winter load: A report on residential customers in Massachusetts prepared by Guidehouse Inc. presents the summer and winter peak day load patterns \cite{Massreport_load_2020}. The summer peak day is a unimodal distribution with a steady increase in electricity usage from 5am to 8pm, and a high afternoon load. In contrast, the winter peak load profile has lower daily electricity consumption and exhibits two modes, with a peak around 8am, and an evening peak around 10pm. The primary difference in summer and winter load is the HVAC (heating, ventilation, and air conditioning) load which increases substatially in the summer months. Notably, the state of Massachusetts does not have widespread electrified heating, which would otherwise result in higher winter loads. These two profiles are adopted here.
\end{itemize}
Note that while data for the Early-Covid19 profile and the Massachusetts peak data are from different geographical regions, both of these regions experience similar climate, with 4 seasons, similar temperatures and precipitation, and similar solar irradiation. As a result, these load profiles are representative of a general distribution grid which may be located in a similar climate. 

\textbf{Mixed load dataset, L2 + L3:} The third load dataset considers a mix of residential loads (as in L2) and commercial loads, which are selected from one of five types. The five commercial loads profiles were selected to be different from one another and different from the residential load profiles of L2. In this way, they introduce new load patterns at different nodes in the grid, and change the net load characteristic as well. The commercial load data is taken from the NREL ComStock\textsuperscript{TM} dataset \cite{NREL_comstock} for the city of Chicago, which occupies Weather Zone 4A, similar to most of Massachusetts. The location of commercial loads was selected by matching the nominal load data from \cite{Su_2005_83dataset} to the peak hourly load from the commercial load data. Some nodes have multiple commercial loads of the same type (ex. multiple retail stores, restaurants, or office buildings at a single node). The location and type of commercial loads are shown in Table~\ref{tab:commercial_load_locations}. Sample data over the first 6 days of January is shown in Fig.\ref{fig:loadprofiles}(c). The five commercial load profiles are described below:
\begin{itemize}
    \item Hospital: The load profile has high temporal characteristics, with higher electricity consumption from 6am to 6pm. The minimum hourly load remains higher than other commercial facilities, at half the peak load. The peak hourly load of the hospital is 1700 kW, corresponding to a hospital of approximately 530,000 sq ft. 
    \item Medium office building: The load profile has a high temporal characteristic with highest load during the morning hours of 5-7am. This commercial building has minimum load during weekends and 10pm-5am. The minimum hourly load is around 20\% of peak load. The peak hourly load is 200 kW.
    \item Quick service restaurant: The load profile has a unique profile in that the load through the day cyclically increases and decreases. There is significant difference in electric load throughout the year, likely depending on customer load through different times of the year. The peak hourly load is 40 kW.
    \item Stand-alone retail space: The load profile for this commercial building is complimentary to the residential load profiles. The load during the day is high for the retail space, corresponding to lower residential loads. The demand begins reducing earlier in the day, approaching the evening residential peak. Similar to the quick service restaurant, there is significant difference in electric load throughout the year. The peak hourly load is 66 kW.
    \item Warehouse: This commercial building has the fewest hours of load from the selected profiles, with sustained high load during the middle of the day. The peak hourly load is 60 kW.
\end{itemize}

\begin{table}
\caption{Locations and type of commercial loads. The labels (1) thru (5) represent commercial loads of the following profiles, respectively: hospital, medium office building, quick service restaurant, stand-alone retail space, and warehouse.}  \label{tab:commercial_load_locations}
    \resizebox{\textwidth}{!}{\begin{tabular}{ll|ll|ll|ll|ll|ll|ll|ll|ll|ll|ll} \\
    \toprule
    \multicolumn{2}{c|}{A} & \multicolumn{2}{c|}{B} & \multicolumn{2}{c|}{C} & \multicolumn{2}{c|}{D} & \multicolumn{2}{c|}{E} & \multicolumn{2}{c|}{F} & \multicolumn{2}{c|}{G} & \multicolumn{2}{c|}{H} & \multicolumn{2}{c|}{I} & \multicolumn{2}{c}{J} & \multicolumn{2}{c}{K} \\ \hline
    $j$ & Label & $j$ & Label & $j$ & Label & $j$ & Label & $j$ & Label & $j$ & Label & $j$ & Label & $j$ & Label & $j$ & Label & $j$ & Label & $j$ & Label\\ \hline
    13 & 4 & 23 & 2 & 28 & 2 & 36 & 5 & 42 & 4 & 55 & 3 & 61 & 2 & 74 & 2 & 77 & 5 & 86 & 3 & 90 & 5\\
    16 & 2 & & & 30 & 1 & 37 & 4 & 47 & 4 &  &  & & & & & 82 & 2 & 87 & 2 & 93 & 4\\
    17 & 1 & & & 34 & 5 & 39 & 1 & 52 & 2 & & & & & & & & & & & & \\
    & & & & 35 & 5 & 40 & 2 & & & & & & & & & & & & & &\\
    \bottomrule
    \end{tabular}}
\end{table}

\subsubsection{Distributed Generation: Solar Dataset}
We add a range of distributed solar facilities to the network, up to a penetration of 23\% of nameplate capacity to baseline load. This is a modest DER penetration, which once again reflects the solar uptake now and over the next few years. The location of solar facilities was selected to encourage both local utilization of generation, as well as exports from one feeder to another. Two sets of solar location and capacity data are created: S1 corresponding to the training data (see Table~\ref{tab:solar_locations_83}) and `Solar Error' corresponding to a set of test data (see Table~\ref{tab:solar_locations_alternate_83}). The location and nameplate capacity of each solar facility is provided in the corresponding tables, with each column representing a feeder of the network. Feeders B, H, and K do not have any solar facilities in the training data, while Feeder K has two solar facilities in the test data. The `Solar Error' dataset emulates the reality that system operators have incomplete information of solar PV location and installed capacity, particularly in regions with lower rates of solar adoption, or new growth in installed solar. The lack of DER visibility is a known concern \cite{EPRI_DERVisibility}. In the `Solar Error' dataset the location and capacity of the PV units may be different than what the operator thought they were (i.e. the training data available). Of note, Feeder G is assumed to have very good information so there are no errors in solar installation data, while the operator has no visibility into the two installations at Feeder K. This dataset can also represent changes in solar adoption over time, and tests the ability of the machine learning algorithm to perform accurate predictions when system conditions change. Compared to the $23\%$ penetration in the S1 dataset used for neural training, the `Solar Error' has a penetration of 26\% of nameplate capacity to baseline load.

\begin{figure}[!h]
    \centering
    \includegraphics[width=0.8\textwidth]{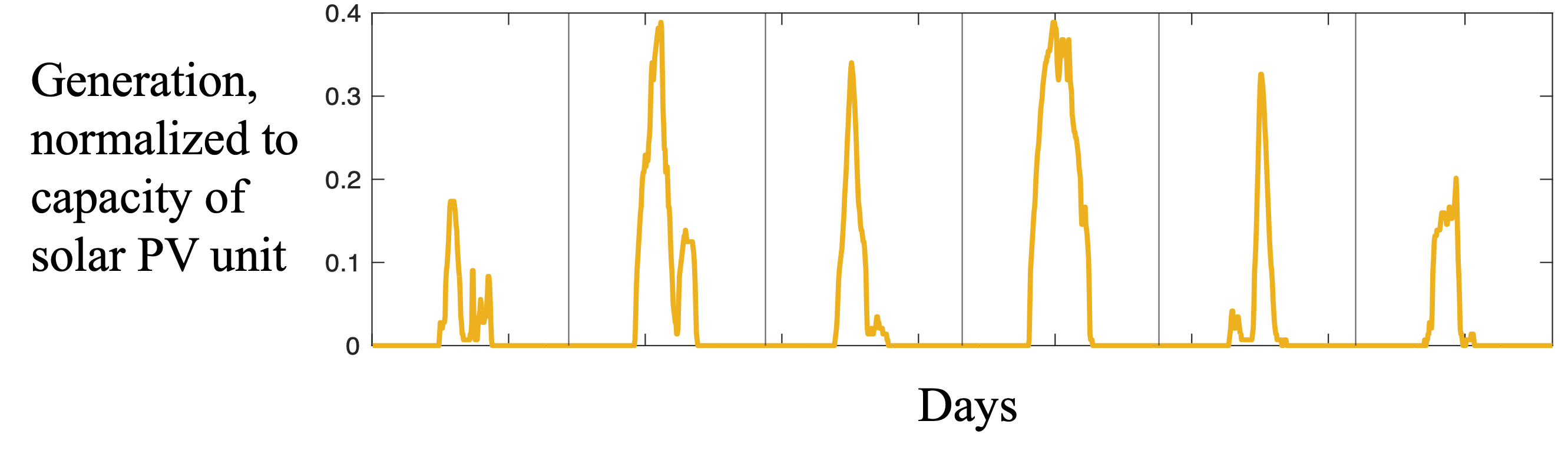}
    \caption{Sample solar PV generation profile for distributed PV located in Massachusetts, across 6 days.}
    \label{fig:genProfile_solarMA}
\end{figure}

To match the load and solar data, the solar generation data is taken from NREL's Solar Power Data for Integration Studies, which consist of synthetic solar PV power plant data points for the United States representing the year 2006. The dataset for Massachusetts was used, for a 12MW distributed PV unit. The dataset provides solar generation at 5-min intervals. The particular file used is \textit{Actual\_42.55\_-72.55\_2006\_DPV\_12MW\_5\_Min.csv}.
Figure~\ref{fig:genProfile_solarMA} shows a sample generation profile of the solar facility, over 6 days.
% https://www.nrel.gov/grid/solar-power-data.html

\begin{table} 
\centering 
\caption{Locations and capacity of the solar facilities per feeder in the TPC-94 grid in training data. The nameplate generating capacity is in kW.}  \label{tab:solar_locations_83} 
    \begin{tabular}{ll|ll|ll|ll|ll|ll|ll|ll}\\
    \toprule
    \multicolumn{2}{c|}{A} & \multicolumn{2}{c|}{C} & \multicolumn{2}{c|}{D} & \multicolumn{2}{c|}{E} & \multicolumn{2}{c|}{F} & \multicolumn{2}{c|}{G} & \multicolumn{2}{c|}{I} & \multicolumn{2}{c}{J} \\ \hline
    $j$ & $\overline{P}^G$ & $j$ & $\overline{P}^G$ & $j$ & $\overline{P}^G$ & $j$ & $\overline{P}^G$ & $j$ & $\overline{P}^G$ & $j$ & $\overline{P}^G$ & $j$ & $\overline{P}^G$ & $j$ & $\overline{P}^G$ \\ \hline
    5 & 150 & 20 & 585 & 40 & 250 & 42 & 450 & 43 & 305 & 48 & 240 & 65 & 180 & 74 & 250 \\
    8 & 200 & 21 & 340 & & & & & 45 & 250 & 50 & 120 & 69 & 340 & & \\
    10 & 330 & 22 & 750 & & & & & 46 & 320 & 53 & 430 & 70 & 250 & & \\
     &  & 24 & 400 & & & & & & & & & & & & \\
    &  & 28 & 350 &  &  &  &  & & & & & & & & \\
    \bottomrule
    \end{tabular}
\end{table}

\begin{table}
\centering 
\caption{Locations and capacity of the solar facilities per feeder in the TPC-94 grid in `Solar Error' test data. The nameplate generating capacity is in kW.}  \label{tab:solar_locations_alternate_83}
    \begin{tabular}{ll|ll|ll|ll|ll|ll|ll|ll|ll}  \\
    \toprule
    \multicolumn{2}{c|}{A} & \multicolumn{2}{c|}{C} & \multicolumn{2}{c|}{D} & \multicolumn{2}{c|}{E} & \multicolumn{2}{c|}{F} & \multicolumn{2}{c|}{G} & \multicolumn{2}{c|}{I} & \multicolumn{2}{c}{J} & \multicolumn{2}{c}{K} \\ \hline
    $j$ & $\overline{P}^G$ & $j$ & $\overline{P}^G$ & $j$ & $\overline{P}^G$ & $j$ & $\overline{P}^G$ & $j$ & $\overline{P}^G$ & $j$ & $\overline{P}^G$ & $j$ & $\overline{P}^G$ & $j$ & $\overline{P}^G$ & $j$ & $\overline{P}^G$ \\ \hline
    3 & 250 & 20 & 585 & 29 & 350 & 36 & 250 & 43 & 325 & 48 & 240 & 65 & 230 & 74 & 250 & 78 & 120\\
    6 & 310 & 21 & 380 & & & 42 & 450 & 45 & 310 & 50 & 120 & 69 & 380 & & & 81 & 150 \\
    8 & 200 & 22 & 750 & & & & & 46 & 290 & 53 & 430 & 70 & 230 & & & & \\
    9 & 280 & 24 & 450 & & & & & & & & & & & & & & \\
    &  & 28 & 350 &  &  &  &  & & & & & & & & & & \\
    \bottomrule
    \end{tabular}
\end{table}

\subsubsection{TPC-94 Datasets}
Using the load and generation data described above, the following five datasets were constructed for neural network training and testing. Each dataset has 17280 data points, generated by using load and solar generation data at 5-minute intervals over 60 days. It must be noted that the load data above includes weekend-weekday variations in demand (most notably for the commercial load profiles), and both load and generation data include seasonal variations (most notably for the solar generation). 
\begin{itemize}
    \item DS1 (Perturbed): Uses perturbed load dataset L1 and solar generation S1, with solar resources introduced in the TPC-94 grid as per Table~\ref{tab:solar_locations_83}
    \item DS2 (Residential): Uses residential load dataset L2 and solar generation S1, with solar resources introduced in the TPC-94 grid as per Table~\ref{tab:solar_locations_83}
    \item DS3 (Mixed): Uses mixed load dataset L2+L3 and solar generation S1, with solar resources introduced in the TPC-94 grid as per Table~\ref{tab:solar_locations_83}
    \item DS4 (Solar error): Uses residential load dataset L2 and solar generation S1, with solar resources introduced in the TPC-94 grid as per Table~\ref{tab:solar_locations_alternate_83}
    \item DS5 (Flat solar): Uses residential load dataset L2, with solar resources introduced in the TPC-94 grid as per Table~\ref{tab:solar_locations_83}. Solar resources are assumed to be generating at nameplate capacity at all times
\end{itemize}

\section{Implications of dynamic reconfiguration on power systems} \label{app:powersystems_results}
We investigate the power systems implications of using Grid-Si\textbf{PhyR} to enable dynamic grid reconfiguration. The following three reconfiguration approaches will be considered:

\textbf{No reconfiguration:} the default grid topology as given in the datasheet is used. All switches are assumed to be in their default position (normally open NOS, or normally closed NCS).

\textbf{Static reconfiguration (StatR):} determines a fixed set of switch-states that optimize losses over a long term, such as a few months or a year. The load and generation forecasts over the period will be considered, and a robust optimization approach can be taken to determine the optimal topology across all grid conditions. 

\textbf{Dynamic reconfiguration (DyR):} determines the switch-states which minimizes losses for a given load and generation condition for a particular period. This period is shorter in length than for StatR, such as a few minutes, hours, or days. The introduction of DERs necessitates DyR where local DERs supply loads in closer proximity to them, thus reducing losses, improving voltage profiles, and increasing PV (renewable soalr energy) utilization. 

The three outcomes (loss reduction, voltage improvement, and increased PV utilization) are investigated below, on simulations of the BW-33 and TPC-94 grids.

\subsection{Power systems performance metrics}
We assess grid performance in terms of grid efficiency, operability, and clean energy directives. The corresponding metrics are as follows:
\begin{itemize}[leftmargin=0.75cm]
    \item \textbf{Grid efficiency:} line losses incurred in delivering power to loads, where power loss across a line $\{i,j\}$ is calculated as $R_{ij}\frac{(P_{ij}-P_{ji})^2 + (Q_{ij}-Q_{ji})^2}{v_i}$ \cite{Baran_reconfig_1989};
    \item \textbf{Grid operability:} voltages must remain within operating limits ($\pm 5$\% of the base system voltage in North America). We consider the voltage distribution, number of undervoltage events ($\sum_{j\in\mathcal{B}}{\mathbb{I}_{\sqrt{V_j} < 0.95}}$), and average grid voltage ($\frac{1}{N}\sum_{j\in\mathcal{B}}{\sqrt{V_j}}$);
    \item \textbf{Clean energy directives:} amount of available PV generation that is dispatched to meet loads. The higher the PV utilization, the more clean energy is used. PV utilization is measured as the amount of PV generation dispatched as a ratio of the available PV resource: $PV_{util} = \frac{P^{G\ast}}{\overline{P}^G}$. %Similarly, the PV curtailment is measured as the amount of PV generation curtailed as a ratio of the available PV resource: $PV_{curt} = \frac{\overline{P}^G - P^{G\ast}}{\overline{P}^G}$. Trivially, $PV_{util} + PV_{curt} = 1$. 
\end{itemize}

\subsection{Result 1: Dynamic reconfiguration reduces electrical line losses}
The following set of results are for the BW-33 grid, a very lossy network: recall that losses average about 8\% of total load.
Table~\ref{tab:reconfig_results} shows the significant loss reduction enabled by dynamic reconfiguration for the BW-33 as DER locations are varied, upwards of 23\% PV penetration. The StatR closes tie line 35 and opens NCS 10, and keeps this topology fixed. The DyR selects primarily between two states: (a) closing tie line 35 and opening NCS 10, and (b) closing tie lines 35 and 36 and opening NCS 10 and 26. For a grid without any DERs versus a grid with DERs, the loss reduction from reconfiguration is higher without DERs; this can be attributed to greater losses without leveraging local generators which are located closer to loads and thus incur lower losses when supplying those loads. The second column compares DyR to StatR, showing savings up to 30 MW for a single distribution feeder. 

While this is a modest 2.5\% improvement of DyR upon StatR, this is a nontrivial reduction for distribution grid operators and utilities. Current industry standard methods which aim to reduce the load on a distribution grid, such as Conservation Voltage Reduction (CVR), typically reduce peak demand by a modest $2$ to $2.5\%$; for a utility with a peak load of 100 MW, this translates to savings of US\$200,000 per year \cite{CVR-1}. Further, it should be noted that this was obtained with a small test case (33 nodes). As the dimension increases, with increasing penetration of DERs and switches, with disparate patterns and topologies, it is expected that this difference may be more pronounced.

\begin{table}[]
\caption{Loss reduction using StatR and DyR}
\label{tab:reconfig_results}
\begin{tabular}{lcc}
\toprule
 & \begin{tabular}[c]{@{}c@{}} StatR vs. no reconfig\\ \% Loss reduction, MW saved per year \textsuperscript{*}\end{tabular} & \begin{tabular}[c]{@{}c@{}}Grid-SiPhyR vs. StatR\\ MW saved per year \textsuperscript{*}\end{tabular} \\ 
 \cmidrule{2-3}
No DERs & 23\%, 370 MW & 0 MW \\
DD-U & 20\%, 300 MW & 23 MW \\
DD-I & 20\%, 320 MW & 31 MW \\
DD-II & 19\%, 270 MW & 20 MW \\
DD-III & 21\%, 310 MW & 27 MW \\
DD-II+III & 20\%, 280 MW & 28 MW \\ \hline
\multicolumn{3}{p{\textwidth}}{*\footnotesize{The MW (power) saved is equal to the MWh (energy), as the simulation is run for every hour of the year and loss reduction summed for every test case}} \\ % \footnotemark
\bottomrule
\end{tabular}
\end{table}

\subsection{Result 2: Dynamic reconfiguration improves voltage profile across the grid}
\begin{figure}[] %!t
	\centering
	\subfloat[No reconfiguration \label{fig:volt_noreconfig}]{\includegraphics[width=0.34\textwidth]{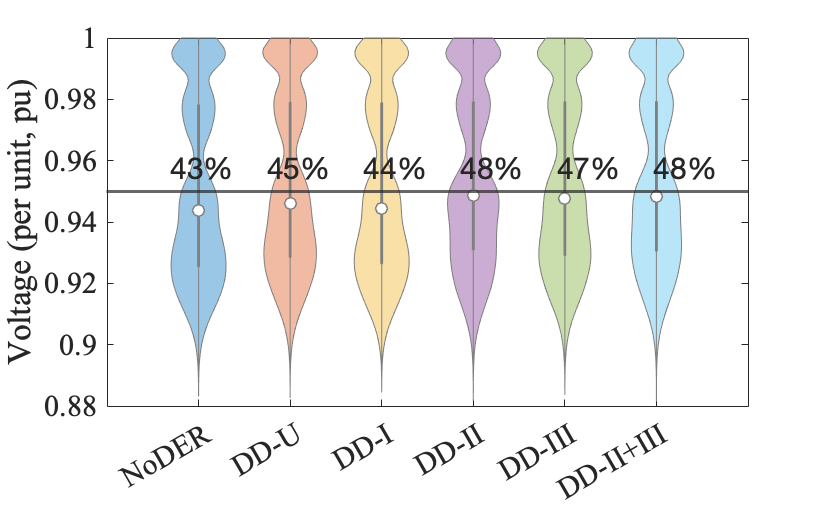}}
% 	\hspace{0.2cm}
	\subfloat[StatR \label{fig:volt_statreconfig}]{\includegraphics[width=0.34\textwidth]{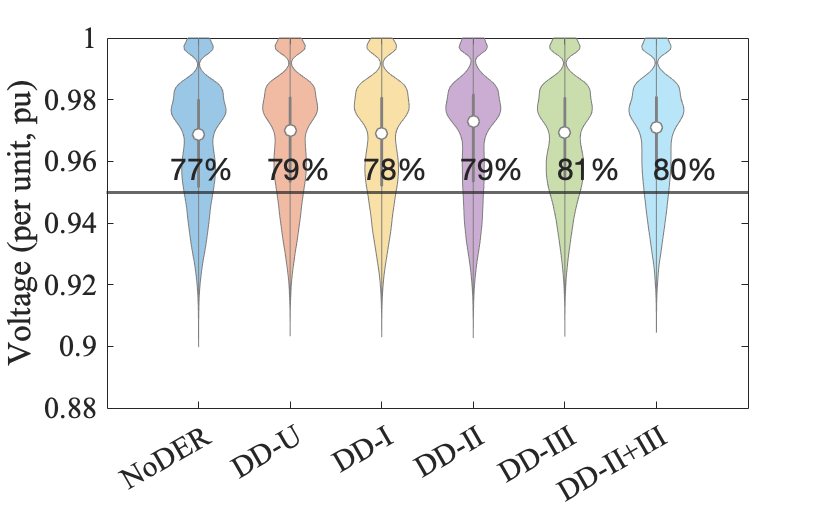}}
% 	\hspace{0.2cm}
	\subfloat[Grid-SiPhyR \label{fig:volt_dynreconfig}]{\includegraphics[width=0.34\textwidth]{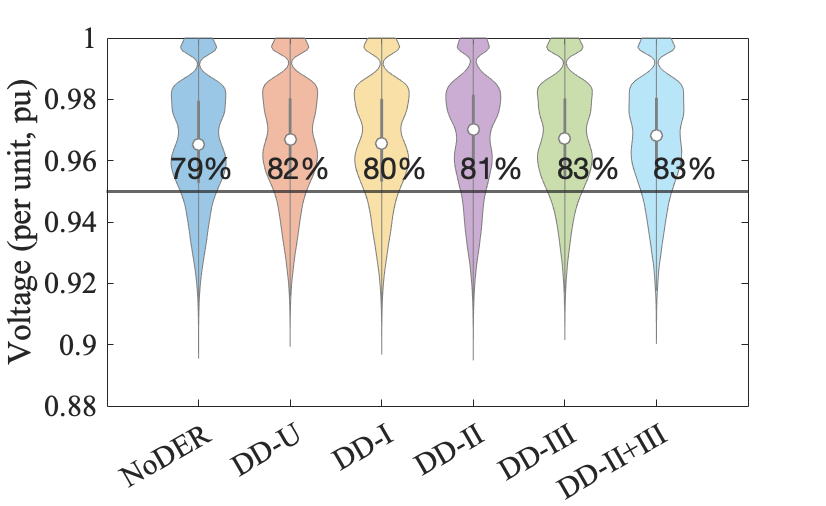}}
	\caption{Voltage distribution over a year (8760 hours). Solid line is ANSI lower voltage limit.}
	\label{fig:results-voltagereconfig}
\end{figure}

The following set of results are for the BW-33 grid, a grid with significant voltage violations. Figure~\ref{fig:results-voltagereconfig} plots the voltage distribution across the grid for an entire year. The ideal distribution is the shape of a short ice cream cone - wide on top and narrow on the bottom, with the tip above the line indicating the ANSI minimum voltage limit of 0.95pu. The width of the plot indicates the total number of voltage observations at the y-axis value, i.e. wider indicates more voltage observations. On each plot is printed the percentage of time the voltages are within the ANSI limits, where higher numbers are better. We make the following key observations: (i) without reconfiguration, the grid performs very poorly, violating ANSI limits 50-60\% of the time, and voltages drop to 0.88pu (outside of ANSI limits); (ii) reconfiguration (Stat or Dy) significantly improves the voltages across the grid, with minimum voltage improving to 0.9pu, and ANSI limits satisfied 77-83\% of the time (IEC limits are always satisfied); (iii) DyR reduces the number of voltage violations throughout the year by 2\%, as compared to StatR, which is a significant improvement as undervoltage can result in brownouts and even lead to blackouts. We note that since the grid chosen for this test case is very lossy, in our simulations we enforce a lower voltage limit of 0.87pu to ensure feasibility of loading conditions (instead of 0.95pu), which our physics-informed framework always satisfies. While it is interesting to note the simple case study of the BW-33 grid does not imply a preferred DD over others, different network topologies and sizes may suggest an optimal DER distribution.
% We anticipate that larger test cases will similarly show greater improvement in the voltage profile with DyR, as with the line losses. 

\subsection{Result 3: Dynamic reconfiguration enables better solar utilization, by connecting generation with loads}
\begin{figure}
    \centering
    \includegraphics[width=\linewidth]{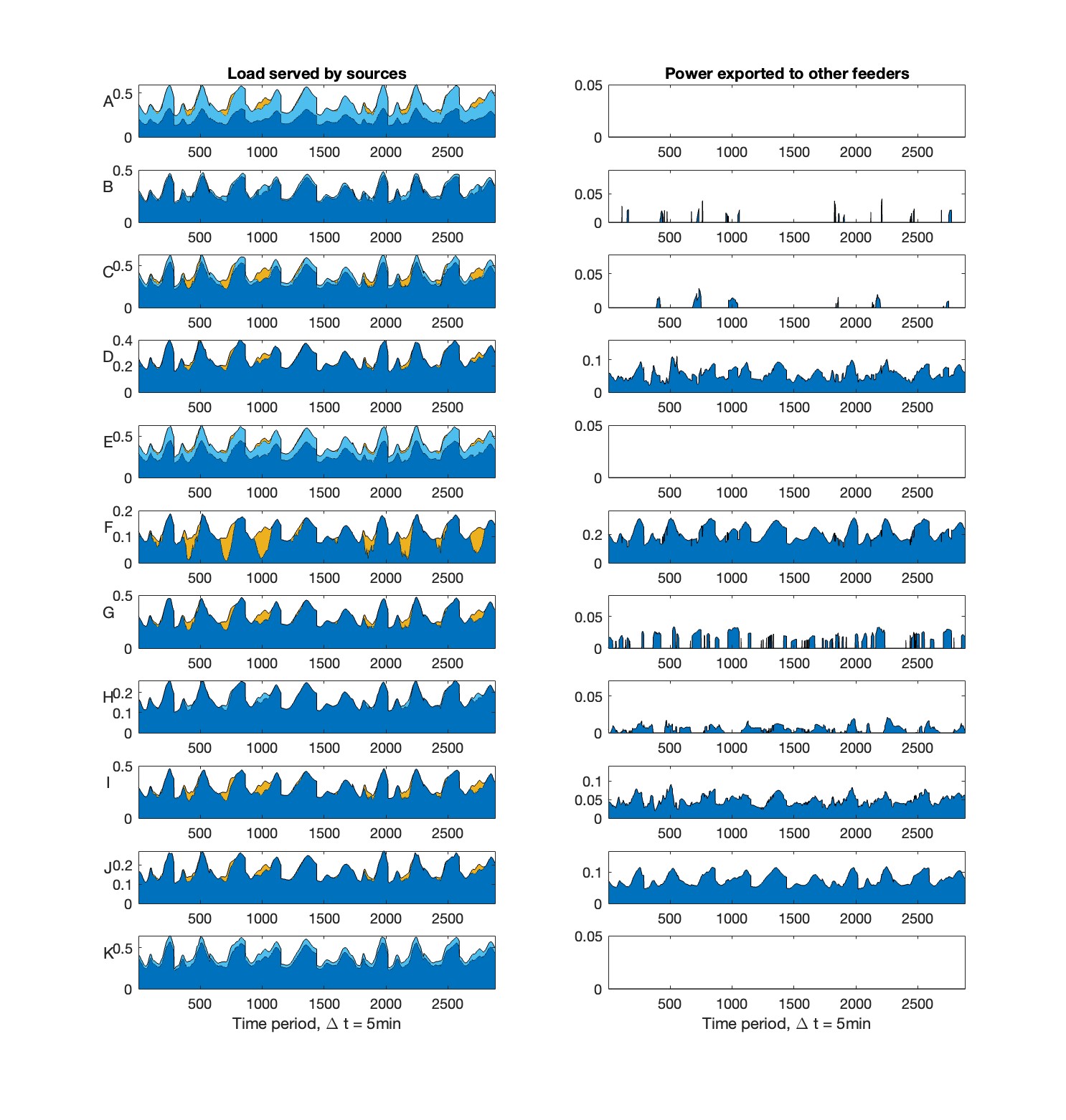}
    \caption{Optimal dispatch results are shown for the TPC-94 network with DS-2 (all residential loads), over a period of 10 days. The optimal reconfiguration is determined every 5-minutes, resulting in 2880 periods. All plots are in per unit measurements. \textbf{(Left)} A breakdown of the load served at each feeder, A thru K. Dark blue: load served by the bulk system at the T-D substation of the same feeder; Light blue: load served by power imported from neighbouring feeders; Yellow: load served by local distributed solar generation. \textbf{(Right)} Power exported to neighbouring feeders.}
    \label{fig:TPC-dispatch-L2}
\end{figure}

The following set of results are for the TPC-94 grid.
Figure~\ref{fig:TPC-dispatch-L2} shows the dispatch results aggregated at the feeder level for the DS2 test case. The load in each feeder can be met by importing power from the bulk grid at the local T-D substation (i.e. the PCC of the same feeder), importing power from neighbouring feeders, or using local PV generation. Most feeders serve the load using imports from the bulk grid, while Feeders A and E import significant amounts of power from neighbouring feeders. Feeder F has very high PV penetration and when solar generation is high, can meet most of its load locally. 

\begin{figure}
    \centering
    \includegraphics[width=\linewidth]{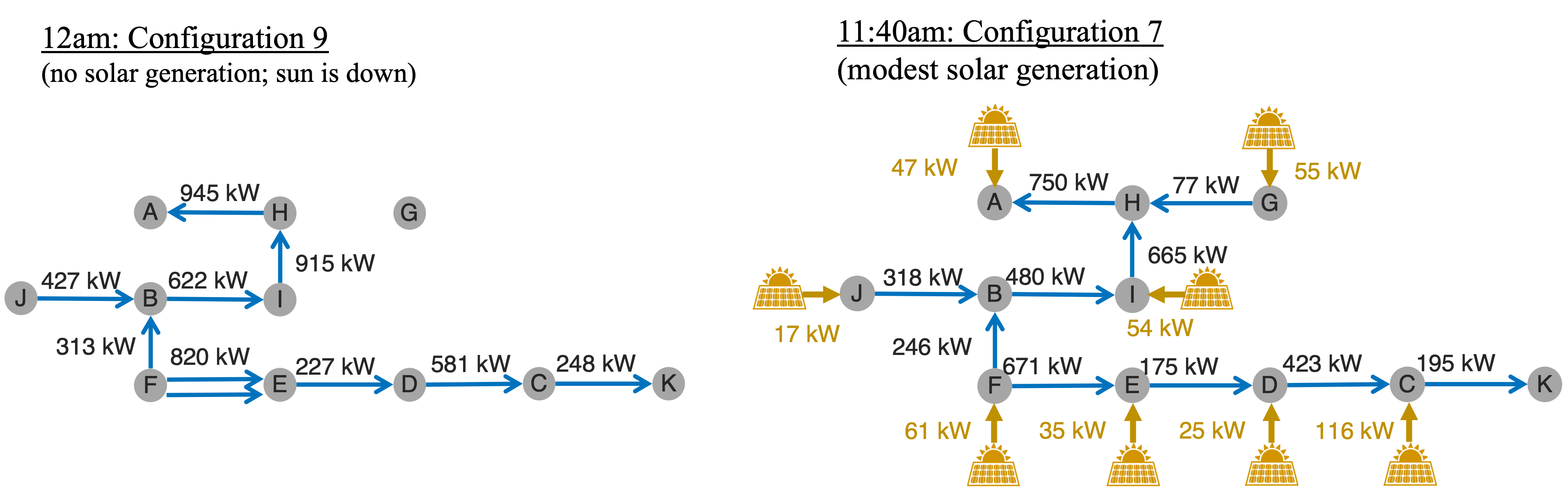}
    \caption{Two configurations for the TPC-94 network showing the switches connecting neighbouring feeders. These results correspond to periods 1 and 140 in the dispatch results of Fig.~\ref{fig:TPC-dispatch-L2}.}
    \label{fig:TPC-configchange-L2}
\end{figure}

Figure~\ref{fig:TPC-configchange-L2} shows a configuration change in the TPC-94 grid, where Feeder G connects to neighbouring Feeder H when solar generation becomes available. Power from Feeder G is exported to Feeder H and A. While the remaining network configuration does not change and the direction of power flow remains the same (i.e. the same networks export power), the amount of power transferred across the tie lines reduces substantially. Local PV generation supplies power to local loads, and the configuration change allows the feeders to meet their remaining load in an efficient way. Notably, the configuration change allows PV utilization to increase in the network.

The PV utilization results are summarized in Table~\ref{tab:summary_PVutil_TPC94}, for StatR and DyR applied to the DS-2 case, and DyR applied to the DS-3 case. The comparison of StatR and DyR show that the optimal reconfiguration of the TPC-94 grid can reduce PV curtailment from 23\% to 19\%. This corresponds to an overall reduction in PV curtailment by 17\%, corresponding to an increase of 250MWh of solar energy used annually. To put into perspective, this is enough solar energy to power approximately 23 US households for a year, and results in a decrease of 107 metric tons of CO\textsubscript{2}.\footnote{This is approximated by 2021 EIA data, where the average annual electricity consumption for a U.S. residential utility customer was 10,632 kilowatthours (kWh), averaging around 886 kWh per month. The CO\textsubscript{2} emissions in Massachusetts from electricity generation in 2021 was 974 lbs/MWh. This value was used to approximate the CO\textsubscript{2} abatement from DyR.} It must be stressed that this increase in solar energy does not require new solar installations. Instead, a change in operating paradigm from static to dynamic permitted by Grid-SiPhyR allows solar PV to be used more effectively, while simultaneously increasing operating efficiency of the distribution grid.

\begin{table}[]
\caption{Summary of PV utilization in TPC-94}
\label{tab:summary_PVutil_TPC94}
\begin{tabular}{llll}
\toprule 
\multicolumn{1}{r}{} & StatR | L2 & DyR | L2 & DyR | L2 + L3\\
\cmidrule{2-4}
Total available PV generation (MWh) & 1042 & 1042 & 880.3 \\
Total utilized PV generation (MWh) & 802.7 & 843.7 & 701.0 \\
Total curtailed PV generation (MWh) & 239.5 & 198.5 & 179.4 \\
Percentage of curtailed PV generation & 23\% & 19\% & 20\%\\
Average utilized PV generation per hour (kWh) & 557.4 & 585.9 & 486.8 \\
Average curtailed PV generation per hour (kWh) & 166.3 & 137.9 & 124.5\\
Number of samples (5-min interval) & \multicolumn{3}{c}{17280} \\
\bottomrule
\end{tabular}
\end{table}

%%%%%%%%%%%%%%%%%%%%%%%%%%%%%%%%%%%%%%%%%%%%%%%%%%%%%%%%%%%%

\end{document}